%% file: RevisionQuaryUpgrade.tex
\newif\ifOneCol

\ifOneCol
\documentclass[article,onecolumn]{IEEEtran}
\else
\documentclass{IEEEtran}
\fi

\usepackage{environ}
\ifOneCol
    
    \newcommand{\singleOnly}[1]{#1}
    \newcommand{\doubleOnly}[1]{}
    \NewEnviron{dcmultline*}{\[ \BODY \]}
    \NewEnviron{dcmultline}{\begin{equation} \BODY \end{equation}}
    \NewEnviron{dcalign}{\begin{equation} \BODY \end{equation}}
\else
    
    \newcommand{\singleOnly}[1]{}
    \newcommand{\doubleOnly}[1]{#1}
    \NewEnviron{dcmultline*}{\begin{multline*} \BODY \end{multline*}}
    \NewEnviron{dcmultline}{\begin{multline} \BODY \end{multline}}
    \NewEnviron{dcalign}{\begin{align} \BODY \end{align}}
\fi

\usepackage{pgfplots}
\usepackage[noadjust]{cite}
\usepackage[utf8]{inputenc} 
\usepackage[T1]{fontenc}
\usepackage{url}
\usepackage{ifthen}
\usepackage{cite}
\usepackage[cmex10]{amsmath} 
\usepackage{amssymb}
\usepackage{epstopdf}
\usepackage{epsfig}
\usepackage{pstricks}
\usepackage{calc,pgf, xcolor}
\usepackage{bbding}
\usepackage{bbm}
\usepackage{dsfont}
\usepackage{ifpdf}
\usepackage{pdfpages}

\newcommand{\Ber}{\mathop{\mathrm{Ber}}}
\newcommand{\Ind}{\mathds{1}}
\newcommand{\calX}{\mathcal{X}}
\newcommand{\calY}{\mathcal{Y}}
\newcommand{\calZ}{\mathcal{Z}}
\newcommand{\Arikan}{Ar\i{}kan}
\newcommand{\myfloor}[1]{\left\lfloor #1 \right\rfloor}
\newcommand{\Pstar}{P^*}
\newcommand{\Prob}{\mathbb{P}}
\newcommand{\m}[1]{\mathcal{#1}}

\newtheorem{theo}{Theorem}
\newtheorem{lemm}[theo]{Lemma}
\newtheorem{claim}[theo]{Claim}
\newtheorem{corollary}[theo]{Corollary}
\newtheorem{definition}[theo]{Definition}
\newtheorem{proposition}[theo]{Proposition}
\newtheorem{remark}[theo]{Remark}

\usepackage{mathtools}
\newcommand{\eqann}[2][=]{\overset{\mathclap{(\text{#2})}}{#1}} 
\newcommand{\eqannref}[1]{$(\text{#1})$}

\newcommand{\twobibs}[2]{#2} 

\begin{document}
\date{}
\author{\IEEEauthorblockN{Or~Ordentlich,~\IEEEmembership{Member,~IEEE}, Ido~Tal,~\IEEEmembership{Senior Member,~IEEE}}
\thanks{Or Ordentlich is with the School of Computer Science and Engineering, Hebrew University of Jerusalem, Israel (email: or.ordentlich@mail.huji.ac.il).  Ido Tal is with the Viterbi Department of Electrical Engineering, Technion, Haifa 32000, Israel (email: idotal@ee.technion.ac.il).}
	\thanks{The work of Or Ordentlich was supported by the ISF under Grant 1791/17. The work of Ido Tal was supported by the BSF under Grant 2018218.
		The material in this paper was presented in part at the 2020 International Zurich Seminar
on Information and Communication~\cite{ot20}.}
	\thanks{}}

\title{An Upgrading Algorithm with Optimal Power Law}
\maketitle
\begin{abstract}
    Consider a channel $W$ along with a given input distribution $P_X$. In certain settings, such as in the construction of polar codes, the output alphabet of $W$ is `too large', and hence we replace $W$ by a channel $Q$ having a smaller output alphabet. We say that $Q$ is upgraded with respect to $W$ if $W$ is obtained from $Q$ by processing its output. In this case, the mutual information $I(P_X,W)$ between the input and output  of $W$ is upper-bounded by the mutual information $I(P_X,Q)$ between the input and output of $Q$. In this paper, we present an algorithm that produces an upgraded channel $Q$ from $W$, as a function of $P_X$ and the required output alphabet size of $Q$, denoted $L$. We show that the difference in mutual informations is `small'. Namely, it is $O(L^{-2/(|\calX|-1)})$, where $|\calX|$ is the size of the input alphabet. This power law of $L$ is optimal. We complement our analysis with numerical experiments which show that the developed algorithm improves upon the existing state-of-the-art algorithms also in non-asymptotic setups.
\end{abstract}
\section{Introduction}

In his seminal paper on polar codes, \Arikan\ introduced  synthetic channels \cite[equation (5)]{Arikan:09p}, also called bit-channels. These synthetic channels have a binary input alphabet and an intractably large output alphabet. Namely, the output alphabet size of such a channel is at least $2^N$, where $N$ is the length of the polar code. When decoding a polar code, this large set does not pose a problem: the decoder must contend with only a single element from the output alphabet. Neither is the output alphabet size a problem when encoding a polar code. However, when \emph{constructing} a polar code, the vast size of the output alphabet is very much an issue. Since polar codes have since been generalized in many ways, let us call the polar codes introduced in \cite{Arikan:09p} `vanilla polar codes'.  To construct a vanilla polar code, one has to pick the `almost noiseless' synthetic channels. That is, essentially, to calculate the probability of a maximum-likelihood (ML) decoder misdecoding the input to the synthetic channel, upon seeing the output of the channel. Clearly, the trivial method of iterating over all possible outputs in order to calculate this quantity will not work, since we cannot iterate over an intractably large set.

A crucial observation is that instead of considering the original synthetic channel, one may approximate it by another channel having a much smaller output alphabet size \cite{MoriTanaka:09c}. Namely, if a channel has a manageable output alphabet size, we may directly calculate the probability of ML misdecoding. A further observation is that if the approximating channel is degraded with respect to the synthetic channel --- the approximating channel can be obtained by processing the output of the synthetic channel --- then the probability of misdecoding in the synthetic channel is upper bounded by the probability of misdecoding in the approximating channel. 

These observations, combined with \cite[Lemma 1.8]{Korada:09z}, were used in \cite{TalVardy:13p} for constructing vanilla polar codes. The key part was an algorithm which transformed a given channel into a degraded approximating channel with a prescribed output alphabet size. The approximating channel was attained by quantizing the output of the synthetic channel to a relatively small alphabet, while decreasing the mutual information between the input and quantized output as little as possible, compared to the mutual information between the input and un-quantized output. This algorithm was used, successively, to approximate each of the $N$ synthetic channels by a corresponding degrading channel. The $k$ approximating channels with the smallest probability of ML misdecoding were used to construct a vanilla polar code with a prescribed dimension $k$. Summing the ML misdecoding probabilities of these $k$ approximating channels gives an upper bound on the probability of misdecoding the polar code using the successive cancellation decoder.

If $W$ is degraded with respect to $Q$, we will also say that $Q$ is upgraded with respect to $W$. In \cite{TalVardy:13p}, a companion algorithm, approximating a given channel by an upgraded channel with a prescribed output alphabet size was also given. In the context of the vanilla polar codes presented in \cite{Arikan:09p}, the importance of this companion algorithm was rather secondary. That is, it resulted in a lower bound on the misdecoding probability of the synthetic channel, and thus one could gauge, through the sandwich property, the closeness of the approximation of the misdecoding probability.

Shortly after their introduction in \cite{Arikan:09p}, polar codes were generalized in various directions. Three important generalizations that soon followed are those to lossy compression \cite{KoradaUrbanke:10p}, asymmetric channels \cite{HondaYamamoto:12p}, and the wiretap channel \cite{HofShamai:10a, MahdavifarVardy:11p,ARTKS:10p,KoyluogluElGamal:12p,rrs13}. In these settings, the construction of the polar code also calls for searching for synthetic channels that are `very noisy'. That is, a maximum a-posteriori (MAP) decoder trying to guess the input to the channel given the output must have a probability of failure close to $1/2$. The same problem of an intractably large output alphabet manifests itself, and the same solution of approximating the channel may be called upon, save for one difference: we now use the upgrading algorithm in \cite{Arikan:09p} in order to lower bound the above probability of MAP misdecoding. Thus, in these contexts, an upgrading approximation is a key part of constructing the polar code, and ceases to be a tool of secondary importance. A more recent example of a polar coding variant in which upgrading plays a key role in the construction process is that of polar codes for settings with memory \cite{Wang+:15c,SasogluTal:19p, ShuvalTal:19.2p, ShuvalTal:18a,TPFV:19a}.

A partial list of papers relating to upgrading and degrading approximations is \cite{PHTT:11c,TalVardy:13c,KochLapidoth:13p,KurkoskiYagi:14p,AlirezaeiMathar:17c,TalSharovVardy:12c,PeregTal:17p,GYB:17p,Tal:17p,KartowskyTal:17p,bnop18,ZhangKurkoski:16c,IwataOzawa:14c,SakaiIwata:17c}. Specifically,  \cite{KartowskyTal:17p} contains a generalization of the upgrading approximation presented in \cite{TalVardy:13p} to cases in which the input has a non-uniform binary distribution. 

The main contribution of this paper is the development and analysis of a novel upgrading algorithm for a general input distribution on an alphabet with cardinality greater than two. Our key idea is to apply a reduction to the binary case, inspired by~\cite{bnop18}, in order to use the algorithm in \cite{KartowskyTal:17p}. Our algorithm will have an optimal power law, a concept we will shortly define.

As a secondary contribution, we show that the channel degrading/quantization algorithm proposed in~\cite[Proof of Proposition 7]{bnop18} for the non-binary input case, has computational complexity which is significantly smaller than the previous state-of-the-art degrading algorithm. Along the way, we also improve the constants in the best known bounds on the degrading and upgrading costs for binary input alphabets under the greedy merge and greedy split algorithms. 

The paper is organized as follows: In Section~\ref{sec:setting} we recall the definition of the channel upgrading problem, while in Section~\ref{sec:MarkovChain} we reformulate it as an optimization problem on Markov chains satisfying several requirements. Our algorithm, along with its performance guarantees is presented in Section~\ref{sec:upgradealg}, and the analysis is carried out in Section~\ref{sec:analysis}. In Section~\ref{sec:degradealg} we turn our attention to the channel degrading/quantization problem and describe a degrading algorithm for a non-binary input alphabet, that attains the optimal power law with complexity that is near-linear in the channel's output cardinality. Finally, In Section~\ref{sec:sim} we demonstrate the practical benefits of our upgrading, as well as degrading, algorithms through numerical experiments. In the analysis of the upgrading and degrading costs of the algorithms we present, we use upper bounds on the upgrading and degrading cost for the case of a binary input. Such bounds with optimal power laws have already appeared in~\cite{KartowskyTal:17p}. In the Appendix we bring somewhat simplified derivations of those bounds, that further result in improved constants.

\section{Setting}
\label{sec:setting}
We are given a channel $W:\calX \to \calY$, where $\calX$ is termed the input alphabet and $\calY$ is termed the output alphabet. We denote the probability of receiving $y \in \calY$ given that $x \in \calX$ was transmitted by $W(y|x)$. We are also given an input distribution $P_X$. That is, we denote by $P_X(x)$ the probability that $x \in \calX$ was the input to the channel. In this paper, we will assume that both $\calX$ and $\calY$ are finite.\footnote{But see the discussion bellow about infinite sized alphabets.}
We denote the mutual information between the input and output of $W$ as
\[
    I(P_X,W) \triangleq I(X;Y) \; , \\
\]
where $X$ and $Y$ are random variables with joint distribution
\begin{equation}
    \label{eq:jointDistribution}
    P_{X,Y}(x,y) = P_X(x) W(y|x) \; .
\end{equation}

Let $Q: \calX \to \calZ$ be a channel with the same input alphabet as $W : \calX \to \calY$. We say that $Q$ is \emph{upgraded} with respect to $W$ if we can obtain $W$ by processing the output of $Q$. That is, if there exists a third channel $\Phi : \calZ \to \calY$ such that, for every $x \in \calX$ and $y \in \calY$,
\[
    W(y|x) = \sum_{z \in \calZ} Q(z|x) \Phi(y|z) \; .
\]

Our goal in this paper is, given an input distribution $P_X$, a channel $W : \calX \to \calY$, and a parameter $L$, to construct a channel $Q:\calX \to \calZ$ that is upgraded with respect to $W$ and whose output alphabet size satisfies $|\calZ| \leq L$. Many such channels $Q$ exist. By the data processing inequality, it must hold that
\[
    I(P_X,Q) \geq I(P_X,W) \; .   
\] 
Hence, our figure of merit of how well $Q$ approximates $W$ will be the difference  $I(P_X,Q) - I(P_X,W)$, which we wish to minimize. It turns out that there exist pairs of input distributions and channels for which such an approximation is inherently `hard'. That is, \cite[Section IV]{KartowskyTal:17p} shows a $P_X$ and $W$ for which\footnote{For sequences $a=a(L)$ and $b=b(L)$, we write $a(L)=O(b(L))$ if there exists constants $C>0$ and $L_0$ such that $a(L)\le Cb(L)$ for all $L \geq L_0$. Similarly, $a(L)=\Omega(b(L))$ means $a(L)\ge Cb(L)$.}
\begin{equation}
    \label{eq:hardChannel}
    I(P_X,Q) - I(P_X,W) = \Omega(L^{-2/(|\calX|-1)}) \; ,
\end{equation}
for every valid choice of $Q$. Our method will produce, for any $P_X$ and $W$, a $Q$ for which 
\begin{equation}
    \label{eq:mutualInformationDifference}
    I(P_X,Q) - I(P_X,W) = O(L^{-2/(|\calX|-1)}) \; .
\end{equation}
Hence, (\ref{eq:hardChannel}) and (\ref{eq:mutualInformationDifference}) imply that our algorithm has an optimal power law. We note that in this paper, $|\calX|$ is assumed to be a fixed constant. That is, in the asymptotic notation used in (\ref{eq:hardChannel}) and (\ref{eq:mutualInformationDifference}), the multiplying constant generally \emph{does} depend on $|\calX|$.

Two comments are in order. First, recall that we have assumed that the output alphabet of $W$ is finite. In many important settings, this will not be the case. For example, if $W$ models the addition of Gaussian noise to a given input supported on a finite alphabet. In such a case, we may use the upgrading algorithm in \cite{PeregTal:17p}, as a preliminary step. That is, the algorithm in \cite{PeregTal:17p} constructs an upgraded channel $Q$ satisfying $I(P_X,Q) - I(P_X,W) = O(L^{-1/(|\calX|-1)})$. Note that this expression is worse than (\ref{eq:mutualInformationDifference}), since the $-2$ in (\ref{eq:mutualInformationDifference}) has been replaced by $-1$. However, for the preliminary step, we may run \cite{PeregTal:17p} with $L^2$ in place of $L$, and then run the algorithm we will shortly introduce on the resulting channel. Clearly, the difference between the original $W$ and the final $Q$ will be $O(L^{-2/(|\calX|-1)})$. Also, it is easy to see that since we have applied two upgrading operations in series, the final channel is upgraded with respect to the initial one. Second, we would like to stress that our algorithm does not generally find the $Q$ which minimizes $I(P_X,Q) - I(P_X,W)$. The problem of finding such a $Q$ has been solved for $|\calX| = 2$ in \cite[Section V]{KartowskyTal:17p}, but remains open for $|\calX| > 2$.

\section{Markov Chain Representation}
\label{sec:MarkovChain}
\subsection{Distributions notation}
In this paper, several probability distributions will be defined and used. For example, we will denote by $\Pstar_{X,Z,Y}(x,z,y)$ a probability distribution on $x \in \calX$, $z \in \calZ$, and $y \in \calY$. The subscript $X,Z,Y$ serves two purposes. The first purpose is to detail how the `corresponding random variables' are defined. Namely, since the first, second, and third entries in the subscript (function arguments list) are $X$ ($x$), $Z$ ($z$), $Y$ ($y$), the corresponding random variables are $X$, $Z$, and $Y$. Also, the probability of $X = x$, $Z = z$ and $Y = y$ equals $\Pstar_{X,Z,Y}(x,z,y)$. We denote the probability of this event as
\[
    \Prob(X = x, Z = z, Y = y) = \Pstar_{X,Z,Y}(x,z,y) \; .
\]
We will always detail under which probability distribution $\Prob$ is calculated. The second purpose of the subscript $X,Z,Y$ is to define derived probability distributions. For example, $\Pstar_{Z|Y}(z|y)$ denotes the probability that $Z = z$ given that $Y=y$, where $Z$ and $Y$ are the random variables described earlier. That is,
\begin{dcmultline*}
        \Pstar_{Z|Y}(z|y) = \Prob(Z=z|Y=y) \\
    = \frac{\sum_{x \in \calX} \Pstar_{X,Z,Y}(x,z,y)}{\sum_{x \in \calX, z' \in \calZ}\Pstar_{X,Z,Y}(x,z',y)} \; .
\end{dcmultline*}
To avoid corner cases (such as division by zero in the above definition), we will always assume without loss of generality (w.l.o.g.) that there are no symbols with probability zero. That is, for each $x \in \calX$ it holds that $\Pstar_{X}(x) > 0$ (otherwise we could remove $x$ from $\calX$), etc.

\subsection{Restatement of setting}
We find it conceptually simpler\footnote{Indeed, one might argue that this is more natural. That is, if we were to coerce the input-distribution/channel terminology to the construction of polar codes for settings with memory \cite{SasogluTal:19p, ShuvalTal:19.2p, ShuvalTal:18a}, using our algorithm for the construction of a polar code would entail defining the `channel input' as consisting of the actual symbol that was fed to the channel, along with a pair of hidden states encapsulating the combined channel and input process state at the beginning and end of the relevant block. Namely, with respect to \cite[equation (25)]{ShuvalTal:19.2p},  the `channel input' is $(U_i,S_0,S_N)$ and the channel output is $Q_i$.} to merge the input distribution $P_X(x)$ and channel $W(y|x)$ into one joint distribution $P_{X,Y}(x,y)$ as in (\ref{eq:jointDistribution}). We will call this the `given distribution'.
We now restate our setting as follows.  \begin{enumerate}
\item We are given a distribution $P_{X,Y}(x,y)$, where $x \in \calX$ and $y \in \calY$, both sets being finite.
\item We must find a distribution $\Pstar_{X,Z,Y}(x,z,y)$, where $x \in \calX$, $z \in \calZ$, $y \in \calY$. We require that $|\calZ| \leq L$.
\item The marginalization of $\Pstar_{X,Z,Y}(x,z,y)$ over $z$ must produce $P_{X,Y}(x,y)$. Namely, for all $x \in \calX$ and $y \in \calY$,
    \[
        P_{X,Y}(x,y) = \sum_{z \in \calZ} \Pstar_{X,Z,Y}(x,z,y) \; .
    \]
\item The corresponding random variables $X,Z,Y$ must form a Markov chain. Namely, we may write, for all $x \in \calX$, $z \in \calZ$, $y \in \calY$,
    \begin{equation}
        \label{eq:MarkovChain}
        \Pstar_{X,Z,Y}(x,z,y) = \Pstar_X(x) \Pstar_{Z|X}(z|x) \Pstar_{Y|Z}(y|z) \; .
    \end{equation}
    That is, the last term $\Pstar_{Y|Z}(y|z)$ is not a function of $x$; our processing of $Z$ to form $Y$ is done without knowledge of $X$.
\item Our figure of merit is the difference
    \begin{dcmultline}
        I(X;Z) - I(X;Y) =  H(X|Y) - H(X|Z) \\
        = I(X;Z|Y) \; , \label{eq:figureOfMerit}
    \end{dcmultline}
    where the second equality follows by the Markov property.
\end{enumerate}

A simple observation that will greatly simplify our derivations is that $X-Z-Y$ form a Markov chain in this order iff $Y-Z-X$ do. That is, we may replace (\ref{eq:MarkovChain}) by the equivalent condition
\begin{equation}
    \label{eq:MarkovChainReversed}
        \Pstar_{X,Z,Y}(x,z,y) = \Pstar_Y(y) \Pstar_{Z|Y}(z|y) \Pstar_{X|Z}(x|z) \; .
    \end{equation}

\section{The algorithm}
\label{sec:upgradealg}

\subsection{Upgradation for the Binary Case}

Our key idea is to apply a reduction from the case in which the input alphabet $\calX$ is non-binary to a case in which the input alphabet is binary. In aid of this, we recall that in \cite[Section VI]{KartowskyTal:17p}, an efficient upgrading algorithm, termed `greedy-split' for the binary-input case is presented. Namely, denote by
\[
    \calX' = \{0,1\}
\]
the binary alphabet. Then, given a joint distribution $P_{X',Y}(x',y)$, where $x' \in \calX'$ and $y \in \calY$, we have an algorithm which produces, for a given $L$, a distribution $\Pstar_{X',Z',Y}(x',z',y)$ such that the conditions in Section~\ref{sec:MarkovChain} are fulfilled,  and the mutual information difference is $O(L^{-2})$.

In Appendices~\ref{subsec:bsp} and~\ref{subsec:binup} we recall the greedy-split algorithm from~\cite{KartowskyTal:17p}, and give a full self contained proof for the bound $I(X';Z')-I(X';Y)=O(L^{-2})$, with improved constants compared to~\cite[Theorem 17]{KartowskyTal:17p}.\footnote{The bound proved in \cite[Theorem 17]{KartowskyTal:17p} was $I(X;Z)-I(X;Y)\leq 2\nu(2) L^{-2}$, where $\nu(2)=64/(\sqrt{3/2}-1)^2$. Thus, we improve the constant by a factor of $(\sqrt{3/2}-1)^{-2}\approx 19.79$.} Namely, we prove the following.\footnote{Throughout, all logarithms are taken to the natural base.}
\begin{theo}
	Let $(X,Y)\sim P_{X,Y}$ be random variables in $\mathcal{X}\times\mathcal{Y}$, where $|\mathcal{X}|=2$ and $\mathcal{Y}$ is discrete. For any natural $L\geq 2$, there exists a random variable $Z$ of cardinality $|\mathcal{Z}|=L$, such that $X-Z-Y$ form a Markov chain in this order, and
	\begin{align}
	I(X;Z)-I(X;Y)\leq 128L^{-2}.
	\label{eq:entropyDifferenceBound}
	\end{align}
	\label{thm:binarysplit}
\end{theo}

Our proof follows that of~\cite[Theorem 17]{KartowskyTal:17p}, essentially step by step, but is simpler and shorter. This is mainly due to a simplification of the ``sphere-packing'' argument. In~\cite{KartowskyTal:17p}, it was argued that given $n$ distributions $P_1,\ldots,P_n$ in the $(q-1)$-dimensional simplex, if $n$ is sufficiently large, we must be able to find $P_i$ and $P_j$, $1\leq i<j\leq n$, such that 
$H(\alpha P_i+(1-\alpha)P_j)-\alpha H(P_i)-(1-\alpha)P_j$ is small,
for all $\alpha\in[0,1]$. The argument leading to this conclusion was of a sphere-packing nature. As the same argument was needed in~\cite{KartowskyTal:17p} also for the analysis of a channel degradation algorithm, termed `greedy-merge', for general alphabet sizes, the sphere-packing argument was derived for general $q$. This led to various technical complications, which in turn led to rather loose constants. Restricting attention to $q=2$, the simplex reduces to the $[0,1]$ interval, and the derivation of the sphere-packing bound is significantly simplified, leading also to better constants in the bound.

\subsection{Upgradation for the General Case}
Denote the input alphabet size as
\[
    |\calX| = q \; .
\]
Similarly to the method in \cite{bnop18}, we will now use the `one-hot' representation of $x \in \calX$ to affect the reduction. Namely, w.l.o.g.\ let us assume that
\[
    \calX = \{1,2,\ldots,q\} \; .
\]
We will replace $x \in \calX$ by a length $q-1$ vector $g(x) = (x_1,x_2,\ldots,x_{q-1})$, such that
\[
    x_i = \begin{cases}
        1 & \mbox{if $x = i$} \\
        0 & \mbox{otherwise}
    \end{cases}
\]
Namely, for $1 \leq i \leq q-1$, we map $x=i$ to the vector $g(x)$ of length $q-1$ that has entry $i$ equal to $1$, and all other entries equal to $0$. We map $x=q$ to $g(q)$, the all-zero vector of length $q-1$. Since the mapping $g$ is invertible, we will often abuse notation and simply write $x = (x_1,x_2,\ldots,x_{q-1})$.

Given the joint distribution $P_{X,Y}$, let $X$ and $Y$ be corresponding random variables. Recalling our convention of denoting $X=(X_1,X_2,\ldots,X_{q-1})$, our first step is to define the following $q-1$ joint distributions: for $1 \leq i \leq q-1$, let
\begin{align}
    \label{eq:jointProbForReduction}
    &\alpha^{(i)}_{X_i,Y}(x',y)  \triangleq  \Prob(X_i = x', Y = y | X_1^{i-1} = 0_1^{i-1} ) \nonumber\\
    &=\Prob(Y=y|X_1^{i-1}=0_1^{i-1})\Prob(X_i = x'| Y = y,X_1^{i-1} = 0_1^{i-1} )\nonumber\\
    &=\alpha^{(i)}_Y(y)\alpha^{(i)}_{X_i|Y}(x'|y),
\end{align}
where $0_1^{i-1}$ is the all-zero vector of length $i-1$, and
\begin{align}
\alpha^{(i)}_Y(y)&\triangleq \Prob(Y=y|X_1^{i-1}=0_1^{i-1})\nonumber\\
\alpha^{(i)}_{X_i|Y}(x'|y)&\triangleq \Prob(X_i = x'| Y = y,X_1^{i-1} = 0_1^{i-1} ).\nonumber
\end{align}
Note that if $i=1$, there is no conditioning. We apply the binary-input upgrading algorithm to each of the above distributions, but require that the resulting output alphabet size be at most
\begin{equation}
    \label{eq:Lambda}
    \Lambda = \myfloor{L^{1/(q-1)}} \; .
\end{equation}
Thus, for each $1 \leq i \leq q-1$, the binary-upgrading algorithm returns a distribution
\[
    \beta^{(i)}_{X_i,Z_i,Y}(x',z',y) \; ,
\]
where $x' \in \calX'$, $z' \in \calZ^{(i)}$ ,$y \in \calY$, and the size of $\calZ^{(i)}$ satisfies $|\calZ^{(i)}| \leq \Lambda$. By definition of upgrading, the induced random variables satisfy the Markov chain $Y-Z_i-X_i$ in this order,  and we may therefore write this distribution as
\begin{IEEEeqnarray*}{rCl}
    \beta^{(i)}_{X_i,Z_i,Y}(x',z',y) &=& \beta^{(i)}_Y(y) \beta^{(i)}_{Z_i|Y}(z'|y) \beta^{(i)}_{X_i|Z_i}(x'|z') \\
                                     &=& \alpha^{(i)}_Y(y) \beta^{(i)}_{Z_i|Y}(z'|y) \beta^{(i)}_{X_i|Z_i}(x'|z') \; , 
\end{IEEEeqnarray*}
and furthermore, we have that the concatenation of the channels $\beta^{(i)}_{Z_i|Y}:\mathcal{Y}\to\mathcal{Z}^{(i)}$ and $\beta^{(i)}_{X_i|Z_i}:\mathcal{Z}^{(i)}\to\mathcal{X}'$ results in the channel $\alpha^{(i)}_{X_i|Y}:\mathcal{Y}\to\mathcal{X}'$. Namely, for all $y\in\mathcal{Y}$ and $x'\in\mathcal{X}'$, we have that
\begin{align}
\alpha^{(i)}_{X_i|Y}(x'|y)=\sum_{z'\in\mathcal{Z}^{(i)}}\beta^{(i)}_{Z_i|Y}(z'|y)\beta^{(i)}_{X_i|Z_i}(x'|z').\label{eq:bindegeq}
\end{align}
We also recall for future use that the corresponding random variables satisfy (\ref{eq:entropyDifferenceBound}), with $\Lambda$ in place of $L$.

We use the above $q-1$ distribution in order to define the distribution $\Pstar_{X,Y,Z}$. Denote $z=(z_1,z_2,\ldots,z_{q-1})$. Also, recall our one-hot convention for $x$, namely $x=(x_1,x_2,\ldots,x_{q-1})$.  Then, for
\[
    \calZ = \calZ^{(1)} \times \calZ^{(2)} \times \cdots \times \calZ^{(q-1)} \; ,
\]
we define for $x \in \calX$, $z \in \calZ$, and $y \in \calY$,
\begin{dcmultline}
    \label{eq:bigDistribution}
    \Pstar_{X,Z,Y}(x,z,y) =P_Y(y) \cdot \left( \prod_{i=1}^{q-1} \beta^{(i)}_{Z_i|Y}(z_i|y) \right) \\
    \cdot \left( \prod_{i=1}^{q-1} \gamma^{(i)}_{X_i|Z_i,X_1^{i-1}}(x_i|z_i,x_1^{i-1})\right) \; ,
\end{dcmultline}
where, for $1 \leq i \leq q$,
\begin{dcmultline}
    \label{eq:gamma}
    \gamma^{(i)}_{X_i|Z_i,X_1^{i-1}}(x_i|z_i,x_1^{i-1}) \\
     = \begin{cases}
                 \beta^{(i)}_{X_i|Z_i}(x_i|z_i) & \mbox{if $x_1^{i-1} = 0_1^{i-1}$} \; , \\
        1 & \mbox{if $x_1^{i-1} \neq 0_1^{i-1}$ and $x_i = 0$} \; , \\
        0 & \mbox{otherwise} \; .
     \end{cases}
\end{dcmultline}

To summarize, our algorithm consists of three steps:
\begin{enumerate}
    \item Construct the distributions $\alpha^{(i)}_{X_i,Y}(x',y)$ for $i=1,\ldots,q-1$, according to ~\eqref{eq:jointProbForReduction}. Each such distribution is supported on $\{0,1\}\times\mathcal{Y}$;
    \item Apply the greedy-split algorithm, as described in~\cite[Section V]{TalVardy:13p} and \cite{KartowskyTal:17p}, to each of the joint distributions $\alpha^{(i)}_{X_i,Y}(x',y)$,  $i=1,\ldots,q-1$, producing the joint distribution $\beta^{(i)}_{X_i,Z_i,Y}(x',z',y)$;
    \item Produce the joint distribution $\Pstar_{X,Z,Y}(x,z,y)$ from the $\beta^{(i)}_{X_i,Z_i,Y}(x',z',y)$, according to (\ref{eq:bigDistribution}) and (\ref{eq:gamma}), and optionally marginalize over $y$ to produce $\Pstar_{X,Z}(x,z)$.
\end{enumerate}

Our main result is the following.
\begin{theo}
The distribution $\Pstar_{X,Z,Y}(x,z,y)$ specified in~\eqref{eq:bigDistribution} is a valid probability distribution, it induces a Markov chain $X-Z-Y$ in this order, and it marginalizes to $\sum_{z\in\calZ}\Pstar_{X,Z,Y}(x,z,y)=P_{X,Y}(x,y)$. Furthermore, under $\Pstar_{X,Z,Y}(x,z,y)$ we have that
\begin{align}
    \label{eq:theoDifferenceInEntropies}
I(X;Z)-I(X;Y)\leq 128(q-1) \myfloor{L^{1/(q-1)}}^{-2}
\end{align}
The construction of $\Pstar_{X,Z,Y}(x,z,y)$, and optionally its marginalization to $\Pstar_{X,Z}(x,z)$ can be done using $O(q|\calY| (\log |\calY|+q\cdot 2^{q-1}))$ simple operations.
\label{thm:main}
\end{theo}

\section{Proof of Theorem~\ref{thm:main}}
\label{sec:analysis}

\subsection{Informal Explanation}

In this subsection we give an intuitive reasoning to why the proposed construction, i.e., the joint distribution on $(X,Z,Y)$ specified by~\eqref{eq:gamma}, indeed induces the correct marginal distribution on $(X,Y)$, satisfies the required Markov relation $X-Z-Y$, and attains a small mutual information gap $I(X;Z)-I(X;Y)$.

\begin{figure*}[t]
    \begin{center}
        \ifpdf
            \includegraphics{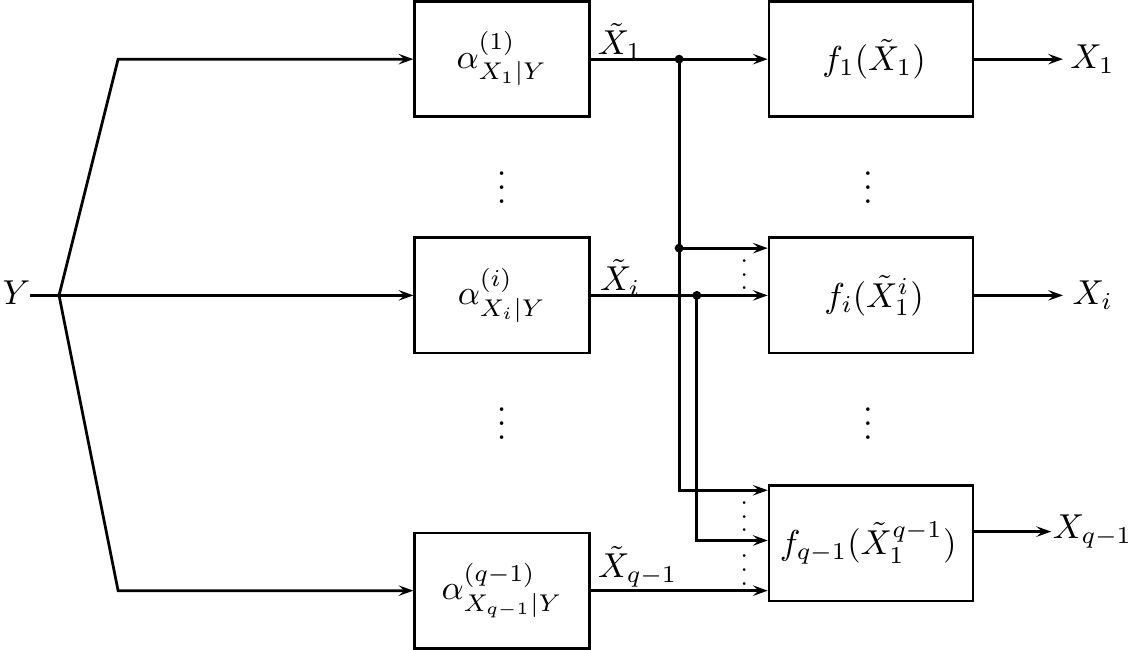}
        \else
            \input{OriginalDistribution_bare.tex}
        \fi
    \end{center}
\caption{A schematic illustration of $P_{YX_1^{q-1}}$, as written in~\eqref{eq:funtionview}. The functions $f_i$ are defined as $f_i(\tilde{x}_1^i)=\tilde{x}_i\cdot \Ind_{\{\tilde{x}_1^{i-1}=0_1^{i-1}\}}$, for $i=1,\ldots,q-1$.}
\label{fig:OriginalDistribution}
\end{figure*}

We begin, by writing the joint distribution on $(X,Y)$ as
\begin{align*}
P_{X,Y}(x,y)&=P_Y(y)P_{X_1,\ldots,X_{q-1}|Y}(x_1,\ldots,x_{q-1}|y)\\
&=P_Y(y)\prod_{i=1}^{q-1}P_{X_i|Y,X_{1}^{i-1}}(x_i|y,x_1^{i-1}).    
\end{align*}
Next, note that by definition of our one-hot encoding, for all $x_1^{i-1}\neq 0_1^{i-1}$ and $y\in\mathcal{Y}$, we have that
\begin{align}
 P_{X_i|Y,X_{1}^{i-1}}(x_i|y,x_1^{i-1})=\begin{cases}
 0 & x_i=1\\
 1 & x_i=0
 \end{cases}.
\end{align}
It follows that we can ``simulate'' the channel $P_{X_i|Y,X_{1}^{i-1}}:\mathcal{Y}\times \{0,1\}^{i-1}\to\{0,1\}$ by first passing $Y$ through the channel $P_{X_i|Y,X_{1}^{i-1}=0_1^{i-1}}:\mathcal{Y} \to\{0,1\}$, and then multiplying the result, which we denote $\tilde{X}_i$, by 
$\Ind_{\{X_1^{i-1}=0_1^{i-1}\}}$. Recalling that by~\eqref{eq:jointProbForReduction}, the channels $P_{X_i|Y,X_{1}^{i-1}=0_1^{i-1}}$ are precisely the channels $\alpha^{(i)}_{X_i|Y}$, and noting that the events $\{X_1^{i-1}=0_1^{i-1}\}$ are equivalent to $\{\tilde{X}_1^{i-1}=0_1^{i-1}\}$, for all $i=1,\ldots,q-1$, we see that 
\begin{dcalign}
    P_{X,Y}(x,y)\doubleOnly{&}=P_Y(y)\doubleOnly{\nonumber}\\
    \cdot\doubleOnly{&}\sum_{\tilde{x}^{q-1}\in\{0,1\}^{q-1}}\prod_{i=1}^{q-1}\alpha^{(i)}_{X_i|Y}(\tilde{x}_i|y)\prod_{i=1}^{q-1}\Ind_{\{x_i=f_i(\tilde{x}_1^{i})\}}, \label{eq:funtionview}
\end{dcalign}
where $f_i(\tilde{x}_1^i)\triangleq \tilde{x}_i\cdot\Ind_{\{\tilde{x}_1^{i-1}=0_1^{i-1}\}}$.
This distribution corresponds to generating $Y\sim P_Y$, then generating $\tilde{X}_1^{q-1}$ by passing $Y$ through the product channel $\prod_{i=1}^{q-1}\alpha^{(i)}_{X_i|Y}$ and then generating $X_1^{q-1}$ by the deterministic transformation
\begin{align*}
(X_1,\ldots,X_{q-1})=\left(f_1(\tilde{X}_1),\ldots,f_{q-1}(\tilde{X}_1^{q-1}) \right).
\end{align*}
This view of the generation process of $(X,Y)$ is illustrated in Figure~\ref{fig:OriginalDistribution}.

\begin{figure*}[t]
    \begin{center}
        \ifpdf
            \includegraphics{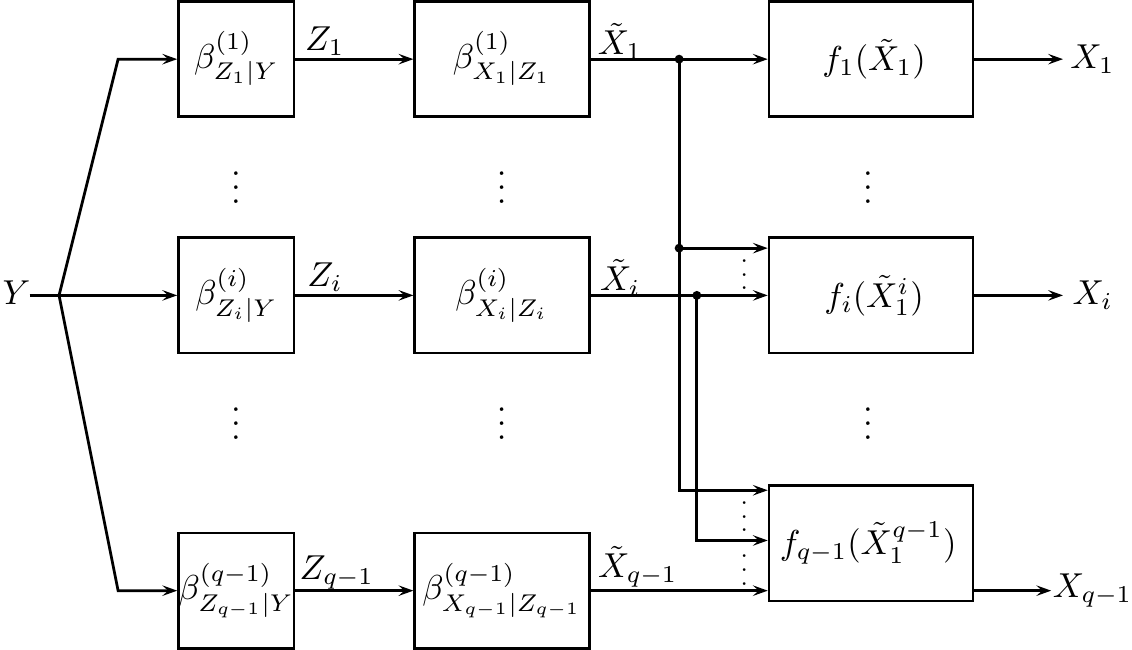}
        \else
            \input{ConstructedDistribution_bare.tex}
        \fi
    \end{center}
\caption{A schematic illustration of the distributed constructed in~\eqref{eq:bigDistribution}. The functions $f_i$ are defined as $f_i(\tilde{x}_1^i)=\tilde{x}_i\cdot \Ind_{\{\tilde{x}_1^{i-1}=0_1^{i-1}\}}$, for $i=1,\ldots,q-1$.}
\label{fig:ConstructedDistribution}
\end{figure*}

Recall that by~\eqref{eq:bindegeq}, for all $i=1,\ldots,q-1$, the channel $\alpha^{(i)}_{X_i|Y}$ is equivalent to the concatenation of the channels $\beta^{(i)}_{Z_i|Y}$ and $\beta^{(i)}_{X_i|Z_i}$. It therefore immediately follows that the joint distribution on $(Y,Z_1^{q-1},X)$ depicted in Figure~\ref{fig:ConstructedDistribution} induces the same marginal distribution on $(X,Y)$ as $P_{X,Y}(x,y)$. Furthermore, under the distribution depicted in Figure~\ref{fig:ConstructedDistribution}, the Markov relation $Y-Z_1^{q-1}-\tilde{X}_1^{q-1}-X$ clearly holds, and consequently, so do the required Markov relation $Y-Z_1^{q-1}-X$. Observing that the distribution depicted in Figure~\ref{fig:ConstructedDistribution} is precisely the one prescribed in~\eqref{eq:bigDistribution}, we conclude that~\eqref{eq:bigDistribution} is indeed a valid distribution for the upgradation problem. 

To quantify the increase of mutual information due to this upgrading procedure, we write
\begin{align}
&I(X;Z)=I(X_1^{q-1};Z_1^{q-1})\nonumber\\
&=\sum_{i=1}^{q-1}I(X_i;Z_1^{q-1}|X_1^{i-1})\nonumber\\
&=\sum_{i=1}^{q-1}\Prob(X_1^{i-1}=0_1^{i-1})I(X_i;Z_1^{q-1}|X_1^{i-1}=0_1^{i-1})\label{eq:only0matters}\\
&=\sum_{i=1}^{q-1}\Prob(X_1^{i-1}=0_1^{i-1})I(\tilde{X}_i;Z_1^{q-1}|X_1^{i-1}=0_1^{i-1})\label{eq:tildeunder0}\; ,
\end{align}
where~\eqref{eq:only0matters} follows since $X_i$ is deterministically equal to $0$ unless $X_1^{i-1}=0_1^{i-1}$, and~\eqref{eq:tildeunder0} follows since $\tilde{X}_i=X_i$ whenever $X_1^{i-1}= 0_1^{i-1}$. Next, we observe that by construction of the distribution, the Markov chain $\tilde{X}_i-Z_i-(Z_{\sim i},X_1^{i-1})$ holds, where $Z_{\sim i} = (Z_1,Z_2,\ldots,Z_{i-1},Z_{i+1},Z_{i+2},\ldots,Z_{q-1})$. This implies that $\tilde{X}_i-(Z_i,X_1^{i-1})-Z_{\sim i}$ form a Markov chain in this order, such that
\begin{align}
I(\tilde{X}_i;Z_{\sim i}|Z_i,X_1^{i-1})=0,\nonumber   
\end{align}
and in particular
\begin{align}
\Prob(X_1^{i-1}=0_1^{i-1})I(\tilde{X}_i;Z_{\sim i}|Z_i,X_1^{i-1}=0_1^{i-1})=0.\label{eq:tildemarkov}
\end{align}
Substituting~\eqref{eq:tildemarkov} into~\eqref{eq:tildeunder0}, we get
\begin{align}
I(X;Z)\leq \sum_{i=1}^{q-1}\Prob(X_1^{i-1}=0_1^{i-1})I(\tilde{X}_i;Z_i|X_1^{i-1}=0_1^{i-1}).\nonumber
\end{align}
Now, recalling that by construction of $\beta^{(i)}_{X_i,Z_i,Y}(x',z',y)$
\begin{IEEEeqnarray*}{rCl}
    I(\tilde{X}_i;Z_i|X_1^{i-1}=0_1^{i-1}) &\leq& I(\tilde{X}_i;Y|X_1^{i-1}=0_1^{i-1})+128\Lambda^{-2} \\
                                           &=& I(X_i;Y|X_1^{i-1}=0_1^{i-1})+128\Lambda^{-2}
\end{IEEEeqnarray*}
and noting that 
\begin{align}
I(X;Y)=\sum_{i=1}^{q-1}\Prob(X_1^{i-1}=0_1^{i-1})I(X_i;Y|X_1^{i-1}=0_1^{i-1}),\nonumber
\end{align}
we obtain
\begin{align}
I(X;Z)-I(X;Y)&\leq \sum_{i=1}^{q-1}\Prob(X_1^{i-1}=0_1^{i-1})128\Lambda^{-2}\nonumber\\
&\leq 128(q-1)\left(\myfloor{L^{1/(q-1)}}\right)^{-2}.\nonumber
\end{align}

\subsection{Formal Proof}
\begin{claim}
    \label{lemm:validProbabilityDistribuion}
    The function $\Pstar_{X,Z,Y}$ defined in (\ref{eq:bigDistribution}) is a valid probability distribution. Specifically, summing the last parenthesized expression in (\ref{eq:bigDistribution}) over all $x \in \calX$ yields $1$; summing the first parenthesized expression in (\ref{eq:bigDistribution}) over all $z \in \calZ$ yields $1$, summing the term $P_Y(y)$ over all $y \in \calY$ yields $1$.
\end{claim}
\begin{IEEEproof}
    By inspection, all the expressions involved are non-negative. Fix some $z \in \calZ$ and consider the last parenthesized expression in (\ref{eq:bigDistribution}). We abuse notation and write
    \begin{multline*}
        \sum_{x \in \calX} \prod_{i=1}^{q-1} \gamma^{(i)}_{X_i|Z_i,X_1^{i-1}}(x_i|z_i,x_1^{i-1})  \\
        =\sum_{(x_1,\ldots,x_{q-1}) \in (\calX')^{q-1}} \; \;  \prod_{i=1}^{q-1} \gamma^{(i)}_{X_i|Z_i,X_1^{i-1}}(x_i|z_i,x_1^{i-1})  \\
        = \prod_{i=1}^{q-1} \sum_{x_i \in \calX'} \gamma^{(i)}_{X_i|Z_i,X_1^{i-1}}(x_i|z_i,x_1^{i-1}) \; ,
    \end{multline*}
    where for the first equality we recall by (\ref{eq:gamma}) that vectors $(x_1,x_2,\ldots,x_{q-1})$ with support greater than $1$ contribute nothing to the above sum, and the RHS is short for 
    \begin{dcmultline*}
        \sum_{x_1 \in \calX'} \gamma^{(1)}_{X_1|Z_1}(x_1|z_1) \sum_{x_2 \in \calX'} \gamma^{(2)}_{X_2|Z_2,X_1}(x_2|z_2,x_1) \\
        \cdots \sum_{x_{q-1} \in \calX'} \gamma^{(q-1)}_{X_{q-1}|Z_{q-1},X_1^{q-2}}(x_{q-1}|z_{q-1},x_1^{q-2}) \; .
    \end{dcmultline*}
Starting from the innermost sum ($i=q-1$) and working out, and recalling the definition of $\gamma^{(i)}$ in (\ref{eq:gamma}), we see that the above equals $1$.

The sub-claim about the first parenthesized quantity in (\ref{eq:bigDistribution}) is proved similarly. The sub-claim about summing $P_Y(y)$ over all $y \in \calY$ follows trivially, by virtue of $P_Y$ being a probability distribution.
\end{IEEEproof}
\begin{claim}
    \label{claim:markovChain} Let the random variables $X$, $Z$, and $Y$ be defined by the distribution $\Pstar_{X,Z,Y}$ given in (\ref{eq:bigDistribution}). Then, $X- Z-Y$ form a Markov chain in this order.
\end{claim}
\begin{IEEEproof}
    We must show that $\Pstar_{X,Z,Y}$ can be factored as in (\ref{eq:MarkovChainReversed}), where each term in this factor is a valid probability distribution. By Claim~\ref{lemm:validProbabilityDistribuion} and inspection of (\ref{eq:bigDistribution}), this is indeed the case.
\end{IEEEproof}
\begin{claim}
    \label{claim:pIsPstarMarginalized}
    Marginalizing the distribution $\Pstar_{X,Z,Y}$ defined in (\ref{eq:bigDistribution}) over all $z \in \calZ$ results in the original joint distribution $P_{X,Y}$.
\end{claim}
\begin{IEEEproof}
    Fix some $x \in \calX$ and $y \in \calY$. We must show that
    \[
 P_{X,Y}(x,y) = \sum_{z \in \calZ} \Pstar_{X,Z,Y}(x,z,y) \; .
    \]
    By inspection of (\ref{eq:bigDistribution}), this is equivalent to proving that
    \begin{dcmultline}
        \label{eq:marginalizationEqualsChannel_zoutside}
        P_{X|Y}(x|y) \\
        = \sum_{z \in \calZ}\prod_{i=1}^{q-1} \beta^{(i)}_{Z_i|Y}(z_i|y) \gamma^{(i)}_{X_i|Z_i,X_1^{i-1}}(x_i|z_i,x_1^{i-1}) \; .
    \end{dcmultline}
    As before, we may simplify the RHS to
    \begin{equation}
        \label{eq:marginalizationEqualsChannel}
        \prod_{i=1}^{q-1} \sum_{z_i \in \calZ_i}\beta^{(i)}_{Z_i|Y}(z_i|y) \gamma^{(i)}_{X_i|Z_i,X_1^{i-1}}(x_i|z_i,x_1^{i-1}) \; .
    \end{equation}
    Recall that $1 \leq x \leq q$. Consider first a term $i$ in the product, where $i > x$. In this case $x_1^{i-1}$ is non-zero, and hence $\gamma^{(i)}_{X_i|Z_i,X_1^{i-1}}(x_i|z_i,x_1^{i-1})$ simply equals $1$, by (\ref{eq:gamma}). Hence, term $i$ in (\ref{eq:marginalizationEqualsChannel}) is simply
    \[
 \sum_{z_i \in \calZ_i}\beta^{(i)}_{Z_i|Y}(z_i|y) = 1 \; .
    \]
    Now consider a term $i$ for in the product (\ref{eq:marginalizationEqualsChannel}) for which $i \leq x$. In this case, $x_1^{i-1}$ is the zero-vector of length $i-1$, and hence $\gamma^{(i)}_{X_i|Z_i,X_1^{i-1}}(x_i|z_i,x_1^{i-1})$ equals $\beta^{(i)}_{X_i|Z_i}(x_i|z_i)$, by (\ref{eq:gamma}). Hence, term $i$ in (\ref{eq:marginalizationEqualsChannel}) is
    \[
        \sum_{z_i \in \calZ_i}\beta^{(i)}_{Z_i|Y}(z_i|y) \beta^{(i)}_{X_i|Z_i}(x_i|z_i)= \alpha^{(i)}(x_i | y) \; ,
    \]
    by virtue of the upgrading process having the marginalization property. Combining these two observations with the definition of $\alpha^{(i)}$ in (\ref{eq:jointProbForReduction}), we simplify (\ref{eq:marginalizationEqualsChannel}) to
    \[
        \prod_{i=1}^{\min\{x,q-1\}} P(X_i = x_i | Y=y, X_1^{i-1}=x_1^{i-1}) \; ,
    \]
    where the probabilities are calculated according to the given probability distribution $P_{X,Y}$.
    Recalling the one-hot convention, we see that for both the case $x=q$ as well as the case $x < q$, the above indeed equals the LHS of (\ref{eq:marginalizationEqualsChannel_zoutside}).
\end{IEEEproof}

\begin{claim}
    \label{claim:upgradingIsEmbeded}
    Let $X$, $Z$, and $Y$ be the random variables corresponding to the distribution $\Pstar_{X,Z,Y}$ defined in (\ref{eq:bigDistribution}). Fix $0 \leq j \leq q-1$. Then, for $x_j \in \calX'$, $z_j \in \calZ_j$, and $y \in \calY$,
          We have
          \begin{dcmultline}
    \label{eq:betaEmbeded}
              \Prob(X_j = x_j, Z_j = z_j, Y = y | X_1^{j-1} = 0_1^{j-1}) \\
              = \beta^{(j)}(x_j,z_j,y) \; .
          \end{dcmultline}
\end{claim}
\begin{IEEEproof}
It suffices to prove that
\begin{dcmultline}
    \label{eq:betaEmbededEquivalent}
    \Prob(X_j = x_j,X_1^{j-1} = 0_1^j, Z_j = z_j, Y = y) \\
    = \beta^{(j)}(x_j,z_j,y) \Prob(X_1^{j-1} = 0_1^{j-1}) \; .
\end{dcmultline}
We get the LHS of (\ref{eq:betaEmbededEquivalent}) by fixing $x_1^{j-1} = 0_1^{i-1}$, and summing $\Pstar(x,z,y)$ over all $x_{j+1}^{q-1}$, $z_{j+1}^q$, and $z_1^{j-1}$. Summing (\ref{eq:bigDistribution}) over only the first two terms, $x_{j+1}^{q-1}$ and $z_{j+1}^q$, causes both products in (\ref{eq:bigDistribution}) to be from $1$ to $j$. Also, since $x_1^{j-1} = 0^{j-1}$, we get from (\ref{eq:gamma}) that the $\gamma^{(i)}$ term in the second product of (\ref{eq:bigDistribution}) can be replaced by $\beta^{(i)}_{X_i|Z_i}(x_i|z_i)$. Thus, we have,
\begin{IEEEeqnarray}{rCl}
    \IEEEeqnarraymulticol{3}{l}{\sum_{x_{j+1}^{q-1}, z_{j+1}^q} \Pstar_{X,Z,Y}(x,z,y)} \IEEEnonumber \\
\quad    &=& P_Y(y) \cdot  \left( \prod_{i=1}^{j} \beta^{(i)}_{Z_i|Y}(z_i|y) \cdot \beta^{(i)}_{X_i|Z_i}(x_i|z_i)\right) \IEEEnonumber \\
         &=& P_Y(y) \cdot  \left( \prod_{i=1}^{j} \beta^{(i)}_{X_i,Z_i|Y}(x_i,z_i|y) \right) \label{eq:towardsBetaEmbeded} \; ,
\end{IEEEeqnarray}
where the second equality follows from the Markovity promised by our upgrading procedure. Recall that we have yet to sum over all $z_1^{j-1}$. Doing so causes all terms on the RHS of (\ref{eq:towardsBetaEmbeded}), save for the term $i=j$ to be marginalized to $\beta^{(i)}_{X_i|Y}(x_i|y)$. Now we recall that by the definition of upgrading, $\beta^{(i)}_{X_i|Y}(x_i|y) = \alpha^{(i)}_{X_i|Y}(x_i|y)$, and by the definition of $\alpha^{(i)}$  in (\ref{eq:jointProbForReduction}) we conclude that
\begin{IEEEeqnarray}{rCl}
    \IEEEeqnarraymulticol{3}{l}{\Prob(X_j = x_j,X_1^{j-1} = 0_1^j, Z_j = z_j, Y = y)} \IEEEnonumber \\
\quad & = &\sum_{x_{j+1}^{q-1}, z_{j+1}^q, z_1^{j-1}} \Pstar_{X,Z,Y}(x,z,y) \IEEEnonumber \\
    \quad    &=& P_Y(y) \cdot  \left( \prod_{i=1}^{j-1} \Prob(X_i = 0|Y=y,X_1^{i-1} = 0_1^{i-1}) \right) \IEEEnonumber \\
    \IEEEeqnarraymulticol{3}{r}{\cdot \beta^{(j)}_{X_j,Z_j|Y}(x_j,z_j|y) \; ,} \label{eq:eitherPorPstar}
\end{IEEEeqnarray}
where the probabilities $\Prob(\cdot)$ above are calculated according to the given probability distribution $P_{X,Y}$, by virtue of this being the probability distribution through which the $\alpha^{(i)}$ are defined. We now note that the RHS of (\ref{eq:eitherPorPstar}) can be simplified to
\[
    \Prob(Y = y, X_1^{j-1}=0_1^{j-1}) \cdot \beta^{(j)}_{X_j,Z_j|Y}(x_j,z_j|y) \; ,
\]
where again, the $\Prob$ are according to the given probability distribution $P_{X,Y}$. Thus,
\begin{IEEEeqnarray*}{rCl}
    \IEEEeqnarraymulticol{3}{l}{\Prob(X_j = x_j,X_1^{j-1} = 0_1^j, Z_j = z_j, Y = y)} \\
    \quad & = & \Prob(Y = y, X_1^{j-1}=0_1^{j-1}) \cdot \beta^{(j)}_{X_j,Z_j|Y}(x_j,z_j|y) \\
    & = & \Prob(X_1^{j-1}=0_1^{j-1}) \\
    \IEEEeqnarraymulticol{3}{r}{\phantom{AAA}\quad \cdot \Prob(Y = y | X_1^{j-1}=0_1^{j-1}) \cdot \beta^{(j)}_{X_j,Z_j|Y}(x_j,z_j|y)} \\
    & = &  \Prob(X_1^{j-1}=0_1^{j-1}) \cdot \alpha^{(j)}_Y(y) \cdot \beta^{(j)}_{X_j,Z_j|Y}(x_j,z_j|y) \\
    & = &  \Prob(X_1^{j-1}=0_1^{j-1}) \cdot \beta^{(j)}_Y(y) \cdot \beta^{(j)}_{X_j,Z_j|Y}(x_j,z_j|y) \\
    & = &  \Prob(X_1^{j-1}=0_1^{j-1}) \cdot \beta^{(j)}_{X_j,Z_j,Y}(x_j,z_j,y) \; ,
\end{IEEEeqnarray*}
proving (\ref{eq:betaEmbededEquivalent}).

\end{IEEEproof}

Now that we have established the $\Pstar_{X,Z,Y}$ is a valid probability distribution, in the sense of upgrading $P_{X,Y}$, we give a bound on the upgrading performance. That is, define the random variables $X$, $Z$, and $Y$ according to $\Pstar_{X,Y,Z}$. We now bound $H(X|Y) - H(X|Z)$ from above.
\begin{lemm}
    \label{lemm:upgradingCost}
    Let a joint distribution $P_{X,Y}$ and a parameter $L$ be given. Construct $\Pstar_{X,Z,Y}$ be as defined in (\ref{eq:bigDistribution}). Let the random variables $X$, $Z$, and $Y$ be defined according to $\Pstar_{X,Z,Y}$. Then, 
    \begin{equation}
        \label{eq:upgradingCost}
        H(X|Y) - H(X|Z) \leq 128 \cdot (|\calX|-1) \cdot \Lambda^{-2} \; .
    \end{equation}
\end{lemm}
\begin{IEEEproof}
    Recall that  our figure of merit is $I(X;Z|Y)$, by (\ref{eq:figureOfMerit}). From the chain rule,
    \begin{IEEEeqnarray*}{rCl}
        \IEEEeqnarraymulticol{3}{l}{I(X;Z|Y)} \\
        \quad & = & I(X_1,X_2,\ldots,X_{q-1};Z|Y)\\
                 & = & \sum_{i=1}^{q-1} I(X_i;Z|Y,X_1^{i-1}) \\
                 & \eqann{a} &  \sum_{i=1}^{q-1} P(X_1^{i-1} = 0_1^{i-1}) \cdot  I(X_i;Z|Y,X_1^{i-1}=0_1^{i-1}) \\
                 & \leq &  \sum_{i=1}^{q-1}  I(X_i;Z|Y,X_1^{i-1}=0_1^{i-1}) \; ,
    \end{IEEEeqnarray*}
    where \eqannref{a} follows from the one-hot representation: if $X_1^{i-1} \neq 0_1^{i-1}$, then $X_i$ is a degenerate random variable always equal to $0$. It would be easy to bound the term $I(X_i;Z|Y,X_1^{i-1}=0_1^{i-1})$, if $Z$ were replace by $Z_i$. Namely, Claim~\ref{claim:upgradingIsEmbeded} and our binary-input upgrading algorithm ensures that\footnote{This is only guaranteed for $\Lambda\geq 2$, but since $128\cdot \Lambda^{-2}>\log{2}$ otherwise, we may assume $\Lambda\geq 2$ without loss of generality.}
    \[
        I(X_i;Z_i|Y,X_1^{i-1}=0_1^{i-1}) \leq  128 \cdot \Lambda^{-2} \; .
    \]
    Thus, our result will be proved once we show that 
    \[
        I(X_i;Z|Y,X_1^{i-1}=0_1^{i-1}) = I(X_i;Z_i|Y,X_1^{i-1}=0_1^{i-1}) \; .
    \]
    By the chain rule, this is equivalent to showing that
    \[
        I(X_i;Z_{\sim i}|Y,Z_i,X_1^{i-1}=0_1^{i-1}) = 0 \; ,
    \]
    where $Z_{\sim i} = (Z_1,Z_2,\ldots,Z_{i-1},Z_{i+1},Z_{i+2},\ldots,Z_{q-1})$, as defined above. That is, we must show that $X_i$ and $Z_{\sim i}$ are independent, when conditioning on an event $A$ of the form 
    \[
        A = \{Y = y, Z_i = z_i, X_1^{i-1} = 0_1^{i-1} \} \; .
    \]
    Hence, fix $y \in \calY$, $x_i \in \calX$, and $z \in \calZ$, and let us show that
    \begin{dcmultline}
        \label{eq:fourTermsProductEquality}
        \Prob(A) \cdot \Prob(A, X_i = x_i, Z_{\sim i} = z_{\sim i}) \\
        = \Prob(A, X_i = x_i) \cdot \Prob(A, Z_{\sim i} = z_{\sim i}) \; .
    \end{dcmultline}
    We now use (\ref{eq:bigDistribution}) and (\ref{eq:gamma}) to write each term of (\ref{eq:fourTermsProductEquality}) explicitly. Namely, one easily gets that 
    \begin{IEEEeqnarray*}{l}
        \Prob(A) = 
        {P_Y(y) \cdot \beta^{(i)}_{Z_i|Y}(z_i|y) \prod_{j=1}^{i-1} \beta^{(j)}_{X_j|Y}(0|y)} \\
        \Prob(A,X_i = x_i, Z_{\sim i} = z_{\sim i}) = \\
    \IEEEeqnarraymulticol{1}{r}{P_Y(y) \cdot \beta^{(i)}_{X_i,Z_i|Y}(x_i,z_i|y) \prod_{j=1}^{i-1} \beta^{(j)}_{X_j|Y}(0|y) \prod_{j=i+1}^{q-1} \beta^{(j)}_{Z_j|Y}(z_j|y)}\\
        \Prob(A,X_i = x_i) = \\
    \IEEEeqnarraymulticol{1}{r}{P_Y(y) \cdot \beta^{(i)}_{X_i,Z_i|Y}(x_i,z_i|y) \prod_{j=1}^{i-1} \beta^{(j)}_{X_j|Y}(0|y)}\\
        \Prob(A, Z_{\sim i} = z_{\sim i}) = \\
    \IEEEeqnarraymulticol{1}{r}{P_Y(y) \cdot \beta^{(i)}_{Z_i|Y}(z_i|y) \prod_{j=1}^{i-1} \beta^{(j)}_{X_j|Y}(0|y) \prod_{j=i+1}^{q-1} \beta^{(j)}_{Z_j|Y}(z_j|y)} \; .
    \end{IEEEeqnarray*}
    Using the above, we easily verify (\ref{eq:fourTermsProductEquality}).

\end{IEEEproof}

\begin{IEEEproof}[Proof of Theorem~\ref{thm:main}]
    By Claims~\ref{claim:markovChain} and \ref{claim:pIsPstarMarginalized}, we have indeed constructed an upgrading of the original distribution. Lemma~\ref{lemm:upgradingCost} ensures that the difference in entropies satisfies (\ref{eq:theoDifferenceInEntropies}), by (\ref{eq:Lambda}) and (\ref{eq:upgradingCost}). Also, the output alphabet size is at most $\Lambda^{q-1} \leq L$, which follows by recalling that $Z=Z_1^{q-1}$ and the definition of $\Lambda$ in (\ref{eq:Lambda}).

    All that remains is to discuss the complexity of our algorithm. The construction of the distributions $\alpha^{(i)}$ given in (\ref{eq:jointProbForReduction}) for  $1 \leq i \leq q-1$  is easy to derive, if for each $y \in \calY$ and $0 \leq i \leq q-1$ we have the probabilities $\Prob(Y=y,X_1^i=0^i)$ at hand. Next, we note that $\Prob(Y=y,X_1^i=0^i)$ is readily computed from $\Prob(Y=y,X_1^{i+1}=0^{i+1})$. Thus, the calculation of all of the $\alpha^{(i)}$ takes time $O(q \cdot |\calY|)$.

    In order to calculate the distributions $\beta^{(i)}$ for $1 \leq i \leq q-1$, the binary upgrading algorithm is run, taking time $O(|\calY| \log |\calY|)$ for each of the $q-1$ invocations. Thus, the time needed to construct the joint distribution $\Pstar_{X,Z,Y}$ is $O( q \cdot |\calY| \log |\calY|)$.

    Let us now discuss the complexity of marginalizing  $\Pstar_{X,Z,Y}$  to produce $\Pstar_{X,Z}$. Clearly, this can be accomplished in time  $O(q \cdot |\calY| \cdot |\calZ| )$. However, we can do better by first noting that even though we produce, for $1 \leq i \leq q-1$, a joint probability $\beta^{(i)}(x',y,z_i)$ involving three variables: $x' \in \calX'$ binary, $y \in \calY$, and $z \in \calZ^{(i)}$, this probability distribution is very sparse. Namely, for each $y \in \calY$ there are at most two $z_i \in \calZ^{(i)}$ such that $\beta^{(i)}(y,z_i) > 0$. This is proved by induction on the number of upgrading steps in the binary upgrading algorithm. The key observation is that a symbol which is removed from the output alphabet due to upgrading is `split' between two \emph{neighboring} symbols, as can be seen in (\ref{eq:upgradingSplit}) below.
    
    As the first step of our efficient marginalization, for each $1 \leq i \leq q-1$ and $y \in \calY$, let us build the subset $\calZ^{(i)}(y)$ of $\calZ^{(i)}$ for which the probability $\beta^{(i)}(y,z_i)$ is positive if and only if $z_i \in \calZ^{(i)}(y)$. As explained earlier, the size of such a set  $\calZ^{(i)}(y)$ is always at most $2$. Also, the time needed to construct these sets is $O(q \cdot |\calY|)$, if the binary upgrading algorithm is modified to retain the relevant information.

    Next, in order to calculate $\Pstar_{X,Z}$, we define an array indexed by $x \in \calX$ and $z = (z^{(1)}, z^{(2)}, \ldots, z^{(q-1)}) \in \calZ$. All entries of the array are initialized to $0$. Then, we have an outermost loop on all $y \in \calY$, a mid-level loop on all $z = (z^{(1)}, z^{(2)}, \ldots, z^{(q-1)})$ such that $z^{(i)} \in \calZ^{(i)}(y)$ for all $1 \leq i \leq q-1$, and an inner-most loop on $x$, going from $1$ up to $q$. The operation carried out in the innermost loop is adding $\Pstar_{X,Z,Y}(x,z,y)$ to entry $(x,y)$ of our table. Clearly, when the calculation finishes, entry $(x,z)$ of our table equals $\Pstar_{X,Z}(x,z)$. Note also that by (\ref{eq:bigDistribution}) and (\ref{eq:gamma}), we can share calculations between different values of $x$, such that time needed for the inner most loop to cycle over all $x \in \calX$ is $O(q)$.
    It follows that the cost of marginalizing $\Pstar_{X,Z,Y}$  to produce $\Pstar_{X,Z}$ can be accomplished in time $O(q \cdot 2^{q-1} \cdot |\calY|)$.
\end{IEEEproof}

\section{An Improved Channel Degrading  Algorithm}
\label{sec:degradealg}

While the focus of this paper is on channel upgrading, we now argue that similar ideas can also be used for channel degrading. Recall that a channel $W$ is degraded with respect to a channel $Q$ if $Q$ is upgraded with respect to $W$.

Although the optimal power law for channel degradation was already established in~\cite{KartowskyTal:17p}, we now provide a channel degrading algorithm that attains the same power law with better (smaller) constants, and more importantly, with reduced complexity compared to that of the algorithm analyzed in~\cite{KartowskyTal:17p}.

The algorithm studied in~\cite{KartowskyTal:17p} is the so-called `greedy-merge' algorithm, which was introduced before under various different names~\cite{TalVardy:13p,st99,GYB:17p}. The algorithm starts with a joint distribution $P_{X,Y}$ supported on $\mathcal{X}\times\mathcal{Y}$ and outputs a joint distribution $P_{X,Z}$ supported on $\mathcal{X}\times\mathcal{Z}$, where $|\mathcal{Z}|=L<|\mathcal{Y}|$, and the channel $P_{Z|X}$ is degraded with respect to $P_{Y|X}$. The algorithm can also output the function $f$ that maps (quantizes) symbols from $\mathcal{Y}$ to $\mathcal{Z}$. The algorithm implements an iterative process that merges two symbols of the output alphabet at each step, hence, reducing the size of the output alphabet by one letter. The two symbols that are merged at each step are those whose merging reduces the mutual information between $X$ and the channel output by the smallest amount. In general, one must enumerate over all possible pairs before deciding which pair to merge, leading to a computational complexity of at least $\Omega(|\mathcal{Y}|^2)$ operations, see~\cite{GYB:17p}. However, in the special case where $|\mathcal{X}|=2$, it can be shown (see~\cite[Theorem 8]{TalVardy:13p} and Remark~\ref{rem:adjacentmerge} in the appendix) that the optimal pair can be found by relabeling the $|\mathcal{Y}|$ output symbols such that $p_1\leq \cdots\leq p_{|\mathcal{Y}|}$, where $p_i=\Prob(X=0|Y=i)$, and then finding the best pair for merging among the pairs $(i,i+1)$, $i=1,\ldots,|\mathcal{Y}|-1$. Thus, in the case where $\mathcal{X}$ is binary, the computational complexity of a single merge is reduced to $O(|\mathcal{Y}|\log |\mathcal{Y}|)$. Further noting that after every merge, only two new pairs of adjacent letters arise, we can save many computations, and implement all $|\mathcal{Y}|-L$ iterations at a total computational cost of $O(|\mathcal{Y}|\log |\mathcal{Y}|)$ operations.

Based on the one-hot representation of $X$, it was shown in~\cite{bnop18} that one can leverage any degrading algorithm with optimal power law for $|\m{X}|=2$ to an optimal-power-law degrading algorithm for general alphabet size. However, an explicit algorithm and an accompanying computational complexity analysis was not given there. Below, we explicitly specify such an algorithm and then analyze its performance.

Let $\Lambda\leq |\mathcal{Y}|$ be some positive integer. Our algorithm degrades the channel $P_{Y|X}:\{1,\ldots,q\}\to \mathcal{Y}$ to the channel $P_{f(Y)|X}:\{1,\ldots,q\}\to\{1,\ldots,L\}$, where $f:\mathcal{Y}\to \{1,\ldots,L\}$ is a deterministic function (quantizer) with cardinality $L\leq\Lambda^{q-1}$. The algorithm consists of three steps:
\begin{enumerate}
    \item\label{it:conddistconst} Construct the distributions $\alpha^{(i)}_{X_i,Y}(x',y)$ for $i=1,\ldots,q-1$, according to ~\eqref{eq:jointProbForReduction}. Each such distribution is supported on $\{0,1\}\times\mathcal{Y}$;
    \item\label{it:bingreedynerge} Apply the greedy-merge algorithm, as described in~\cite[Algorithm C]{TalVardy:13p} and \cite{KartowskyTal:17p}, for each of the joint distributions $\alpha^{(i)}_{X_i,Y}(x',y)$,  $i=1,\ldots,q-1$, to degrade the channel $\alpha^{(i)}_{Y|X_i}(y|x')=\mathbb{P}(Y=y|X_i=x',X_1^{i-1}=0_1^{i-1})$ to the channel $P_{f_i(Y)|X_i}$ with $f_i:\mathcal{Y}\to \{1,\ldots,\Lambda\}$.
    \item \label{it:channelaggreg} Set $f:\mathcal{Y}\to\{1,\ldots,\Lambda\}^{q-1}$ as $f(y)=[f_1(y),\ldots,f_{q-1}(y)]$ and compute the resulting joint distribtution $P_{X,f(Y)}$ as
    \begin{align*}
    P_{X,f(Y)}(x,t_1^{q-1})=\sum_{y\in\mathcal{Y}} P_{X,Y}(x,y)\prod_{i=1}^{q-1}\Ind_{\{f_i(y)=t_i\}},    
    \end{align*}
    for all $(x,t_1^{q-1})\in \{1,\ldots,q\} \times \{1,\ldots,\Lambda\}^{q-1}$.
\end{enumerate}

\begin{theo}
Let $(X,Y)\sim P_{X,Y}$ be random variables in $\mathcal{X}\times\mathcal{Y}$, where $|\mathcal{X}|=q>1$, and $\mathcal{Y}$ is discrete. For any integer $L$, the algorithm defined above with $\Lambda=\lfloor L^{1/(q-1)}\rfloor$ satisfies
\begin{align}
I(X;Y)-I(X;f(Y))\leq 64(q-1) \cdot \left\lfloor L^{1/(q-1)}\right\rfloor^{-2}.
\label{eq:degradeonehotcost}
\end{align}
The construction of the joint distribution $P_{X,f(Y)}$ can be done using $O(q|\mathcal{Y}|\log|\mathcal{Y}|))$ simple operations.
\label{thm:greedymergegen}
\end{theo}

\begin{IEEEproof}
The proof of~\eqref{eq:degradeonehotcost} is similar to~\cite[Proof of Proposition 7]{bnop18}, albeit with improved constants. The improved constants result from a tighter analysis of the degradation loss of the greedy-merge algorithm for the binary case $|\mathcal{X}|=2$, which is derived in Appendix~\ref{subsec:bindown}. In Appendix~\ref{subsec:gendown} we prove~\eqref{eq:degradeonehotcost} based on this result.

As for the time complexity: Step~\ref{it:conddistconst} was shown to be performed using $O(q|\mathcal{Y}|)$ simple operations in the proof of Theorem~\ref{thm:main}. Each application of the greedy-merge algorithm for a binary input alphabet can be performed in $O(|\mathcal{Y}|\log |\mathcal{Y}|)$ operations~\cite{TalVardy:13p} and therefore the computational cost of step~\ref{it:bingreedynerge} is $O(q|\mathcal{Y}|\log |\mathcal{Y}|)$. 
Finally, for the computation of Step~\ref{it:channelaggreg}, we can construct a table of size $q \times \Lambda^{q-1}$ and initialize it with zeros. Then, for each $y\in\mathcal{Y}$, we first calculate $f(y)= [f_1(y),\ldots,f_{q-1}(y)]$, which can be done in time $O(q)$ for each $y$, since each of the $f_i$ can be implemented using an array (lookup table). Thus, we have calculated which column to update in our table. We now loop over all $x$ and increase the value of the cell indexed by $(x,f(y))$ by $P_{X,Y}(x,y)$. Once we have gone through all $y\in\mathcal{Y}$ the table holds $P_{X,f(Y)}$. Thus, Step~\ref{it:channelaggreg} can be performed using $O(q |\mathcal{Y}|)$ simple operations.
\end{IEEEproof}



\section{Numerical Results}
\label{sec:sim}
We now discuss some numerical results that demonstrate the effectiveness of our method. First, consider Figure~\ref{fig:loglog_upgrade}. In it, we fix the input alphabet size to $q=3$ and give four plots of $\Delta I$, the difference in mutual information resulting from upgrading, as a function of $L$, the output alphabet size of the upgraded channel. The topmost plot is the upper bound (\ref{eq:theoDifferenceInEntropies}) we have derived for our one-hot upgrading algorithm. The second and third plots from the top are the result of running the upgrading algorithm in \cite{PeregTal:17p} and our one-hot upgrading algorithm, respectively. The two algorithms were run on the channel described in \cite[Subsection V.A]{Tal:17p}, with parameter $M=400$ and uniform input distribution. As discussed in \cite[Section IV]{KartowskyTal:17p}, this channel is ``hard to upgrade''. We stress that for these two plots, the $x$-axis is in fact $L'$, the number of output letters in the resulting upgraded channel, \emph{not} the ``designed'' output alphabet size $L$ which is the input to these algorithms. The bottom-most plot is the lower bound on the Upgrading Cost, given in \cite[equation (43)]{KartowskyTal:17p}. Fist, note that all plots save the second from the top have roughly the same slope. This is because the two bounds have the same power law, $L^{-2/(q-1)}$, and we have chosen to upgrade a ``hard'' channel for which this power law is mandated, and indeed achieved by our algorithm. In contrast, we can see that the slope corresponding to the upgrading algorithm in \cite{PeregTal:17p} is roughly half of that as ours. This is explained by noting that \cite{PeregTal:17p} only promises a power law of $L^{-1/(q-1)}$.

\begin{figure}
\begin{center}
	\input{numerical_example_upgrade.tex}
	\end{center}
    \caption{\textbf{Upgrading}. Log-log plot of the difference in  mutual information as a function of output alphabet size. For input alphabet size $q=3$, we plot, from top to bottom: the upper bound in (\ref{eq:theoDifferenceInEntropies}); the algorithm from \cite{PeregTal:17p}; our one-hot upgrading algorithm; the lower bound on Upgrading Cost given in \cite[equation (43)]{KartowskyTal:17p}.}
    \label{fig:loglog_upgrade}
    
\end{figure}

Figure~\ref{fig:loglog_degrade} is the analog of Figure~\ref{fig:loglog_upgrade}, but now we consider degrading. Again, we have four plots. From top to bottom, these are: the upper bound in (\ref{eq:degradeonehotcost}) on the one-hot degrading algorithm; the result of running the algorithm in \cite{TalSharovVardy:12c} on the same channel as before (which \cite{Tal:17p} proves is ``hard to degrade''); running the one-hot degrading algorithm (originally presented in \cite{bnop18}, and discussed in our Section~\ref{sec:degradealg}) on this channel; the lower bound on the Degrading Cost given in \cite[equation (3)]{Tal:17p}. Note that the one-hot degrading algorithm is indeed better than the prior art in \cite{TalSharovVardy:12c}, but this is not as pronounced as it was in the upgrading case. We conjecture that this is because the power law of the algorithm in \cite{TalSharovVardy:12c} can be strengthened to something better than $L^{-1/(q-1)}$ (proven in \cite[Lemma 8]{PeregTal:17p}). 

\begin{figure}
\begin{center}
	\input{numerical_example_degrade.tex}
	\end{center}
    \caption{\textbf{Degrading}. Log-log plot of the difference in  mutual information as a function of output alphabet size. For input alphabet size $q=3$, we plot, from top to bottom: the upper bound in (\ref{eq:degradeonehotcost}); the algorithm from \cite{TalSharovVardy:12c}; the one-hot degrading algorithm; the lower bound on Degrading Cost given in \cite[equation (3)]{Tal:17p}.}
    \label{fig:loglog_degrade}
\end{figure}

We close this section with another comparison between our upgrading algorithm and that of \cite{PeregTal:17p}. As explained in the introduction, the upgrading transform is used in order to construct polar codes whose input distribution is non-symmetric. In light of this, consider Table~\ref{tbl:upgrading}. In it, each column corresponds to a different input distribution on the symbols of $\calX = \{0,1,2\}$. That is the second column corresponds to an input distribution in which the symbol $0$ has an a priori probability of $0.8$, and the two symbols $1$ and $2$ each have an a priori probability of $0.1$. The second row in the table is the entropy corresponding to each input distribution. The rest of the table is divided into two parts: the first part corresponds to the one-hot upgrading algorithm, and the second part to the upgrading algorithm from \cite{PeregTal:17p}. For both algorithms, the designed output alphabet size is $100$. That is, for the one-hot algorithm we take $L=100$ and for the algorithm in \cite{PeregTal:17p} we take $\mu$ such that $[0,1]$ is comprised of $10$ regions. In both parts, we consider building a code of length $256$, gotten by applying $8$ polar transforms. The first row in each block is the average entropy of $H(U_i|U^{i-1})$, where $1 \leq i \leq 256$. Ideally, this average entropy should equal that of underlying input distribution. However, do to the quantization effect of upgrading, the average entropy is lower. We see, again, that the one-hot algorithm results in a much smaller reduction of entropy. The next two rows count the number of indices $i$ for which the total variation parameter $K(U_i|U^{i-1})$ is smaller than a prescribed number, which is $0.01$ for the penultimate row in each part and $0.001$ for the last row in each part. The quantity $K(U_i|U^{i-1})$ is as defined in \cite[Appendix B]{ShuvalTal:19.2p}. We want this number to be as high as possible. That is, we count the pool of $U_i$ with corresponding total-variation low enough to allow them to be ``non-frozen'' (and from this pool, we will only choose those whose Bhattacharyya parameter $Z(U_i|U^{i-1},Y^N)$ is low as well, in order to facilitate decoding). Again, one-hot decoding performs substantially better than \cite{PeregTal:17p}.

\begin{table}
    \begin{center}
    \caption{Comparison of one-hot upgrading and \cite{PeregTal:17p}.}
    \setlength{\tabcolsep}{2pt}
    \begin{tabular}{|c|c|c|c|c|}
        \hline
        $(p_0,p_1,p_2)$ & $(0.8,0.1,0.1)$ & $(0.6,0.2,0.2)$ & $(0.4,0.3,0.3)$ & $(0.34,0.33,0.33)$ \\ \hline
     $H$ & $0.92192$ & 1.370951 & 1.57095 & 1.58482 \\ \hline \hline
     one-hot & & & & \\
     avg. $H$ & 0.89484 & 1.34842 & 1.56692 & 1.58473 \\
     $@0.01$ & $68$ & $150$ & 231 & 253 \\
     $@0.001$ & 50 & 129 & 216 & 247  \\ \hline \hline
     \cite{PeregTal:17p} & & & & \\
     avg. $H$ & 0.50171 & 0.75992 & 1.07587 & 1.33681 \\
     $@0.01$ & 8 & 21 & 63 & 116 \\
     $@0.001$ & 8 & 16 & 33 & 65 \\ \hline \hline
    \end{tabular}
    \label{tbl:upgrading}
\end{center}
\end{table}

A reader wishing to reproduce the above results (or obtain new ones) can use the code we have published on \cite{Tal:21g}. 

\begin{appendices}
\section*{Appendix: Upgradation and Degradation in the Binary Case}\label{app:binarymerge}

The purpose of this appendix is to sharpen some of the results of~\cite{KartowskyTal:17p}, as well as provide simpler and shorter proofs. The main contribution of the appendix is a simple derivation of a sphere-packing bound for the simple case of $|\mathcal{X}|=2$, provided in Appendix~\ref{subsec:bsp}. Then, in Appendix~\ref{subsec:binup} and Appendix~\ref{subsec:bindown}, we essentially repeat the arguments from~\cite{KartowskyTal:17p} in order to leverage the improved sphere-packing bound of Appendix~\ref{subsec:bsp} to improved upper bounds on the loss of upgradation and degradation, respectively, for the case of $|\mathcal{X}|=2$. While these bounds are sharper than those reported in~\cite{KartowskyTal:17p}, the proofs in these subsections do not contain new ideas and are brought merely for completeness. As this paper shows, upper bounds on the loss of upgradation for this special case, immediately yield tight bounds (in terms of the power-law) for general $|\mathcal{X}|$. Similarly, in~\cite[Proposition 7]{bnop18} it was shown that upper bounds on the loss of degradation in the binary case yields tight power-law bounds for general $|\mathcal{X}|$. In Appendix~\ref{subsec:gendown} we repeat that derivation in order to prove Theorem~\ref{thm:greedymergegen} from Section~\ref{sec:degradealg} and obtain sharper bounds on the loss for degradation for general $|\mathcal{X}|$.

\section{Sphere-Packing in the Simplex of Bernoulli Distributions}
\label{subsec:bsp}

Let $h_2(p)=-p\log(p)-(1-p)\log(1-p)$ be the binary entropy function. By the concavity of $p\mapsto h_2(p)$ we have that $h_2(\alpha p_0+(1-\alpha)p_1)-\alpha h_2(p_0)-(1-\alpha)h_2(p_1)\geq 0$ for all $\alpha,p_0,p_1\in[0,1]$. The next lemma upper bounds this difference universally for all $\alpha\in[0,1]$, using relatively simple functions of $p_0,p_1$. We remind the reader that logarithms are taken to the natural base.

\begin{lemm}
For any $0\leq p_0\leq p_1\leq1$ and $\alpha\in[0,1]$, we have that
\begin{dcalign}
    h_2(\alpha p_0\doubleOnly{&}+(1-\alpha)p_1)-\alpha h_2(p_0)-(1-\alpha)h_2(p_1)\nonumber\\
    \doubleOnly{&}\leq \min\left\{p_1-p_0,\; \frac{(p_1-p_0)^2}{2\min\{p_0,1-p_1\}}\right\}.\nonumber
\end{dcalign}
\label{lem:entconcbound}
\end{lemm}

We note that, up to constants, the upper bound $p_1-p_0$ and the upper bound $\frac{(p_1-p_0)^2}{2\min\{p_0,1-p_1\}}$, respectively, can be obtained by specializing~\cite[Lemma 1]{GYB:17p} and~\cite[Equation (11-12)]{KartowskyTal:17p} to the binary case.

\begin{IEEEproof}
Let $A\sim\Ber(1-\alpha)$ and let $B$ be a binary random variable with conditional distributions $[B|A=0]\sim\Ber(p_0)$, and $[B|A=1]\sim \Ber(p_1)$. With these definitions, we have that
\begin{align}
h_2(\alpha p_0+(1-\alpha)p_1)-\alpha h_2(p_0)-(1-\alpha)h_2(p_1)=I(A;B).\nonumber
\end{align}
Using the variational formula for mutual information~\cite[Chapter 2, Equation (3.7)]{ck81}, we write
\begin{align}
I(A;B)&= \min_Q D(P_{B|A}\|Q|P_A)\nonumber\\
&=\min_p \alpha\cdot d_2(p_0\|p)+(1-\alpha)\cdot d_2(p_1\|p),\label{eq:dp}
\end{align}
where $D(P\|Q)$ is the KL divergence between $P$ and $Q$, $D(P_{X|A}\|Q_{X|A}|P_A)=\mathbb{E}_{a\sim P_A}[D(P_{X|A=a}\|Q_{X|A=a})]$, and $d_2(p\|q)=p\log\frac{p}{q}+(1-p)\log\frac{1-p}{1-q}$ is the binary KL divergence. To obtain an upper bound, we may take $p=\frac{p_1}{1+p_1-p_0}$ in~\eqref{eq:dp}. Noting that this choice satisfies $p_0\leq p\leq p_1$, we have that
\begin{align}
d_2(p_0\|p)&=p_0\log\frac{p_0}{p}+(1-p_0)\log\frac{1-p_0}{1-p}\nonumber\\
&\leq (1-p_0)\log\frac{1-p_0}{1-p}\nonumber\\
&=(1-p_0)\log(1+p_1-p_0),\label{eq:dp1}
\end{align}
and
\begin{align}
d_2(p_1\|p)&=p_1\log\frac{p_1}{p}+(1-p_1)\log\frac{1-p_1}{1-p}\nonumber\\
&\leq p_1\log\frac{p_1}{p}\nonumber\\
&=p_1\log(1+p_1-p_0),\label{eq:dp2}
\end{align}
Substituting~\eqref{eq:dp1} and~\eqref{eq:dp2} into~\eqref{eq:dp}, yields
\begin{dcalign}
    h_2(\alpha p_0+(1-\alpha)p_1)\doubleOnly{&}-\alpha h_2(p_0)-(1-\alpha)h_2(p_1)\doubleOnly{\nonumber}\\
    \doubleOnly{&}\leq \log(1+p_1-p_0)\leq p_1-p_0.\label{eq:ub1}
\end{dcalign}
We will obtain another upper bound using the fact that KL divergence is dominated by $\chi^2$ divergence. Specifically~\cite[eq. (5)]{SasonVerdu:16}, $D(P\|Q)\leq\log(1+\chi^2(P\|Q))$. For the binary case, this bound reads
\begin{align}
d_2(p\|q)\leq\log\left(1+\frac{(p-q)^2}{q(1-q)}\right)\leq \frac{(p-q)^2}{q(1-q)}.
\end{align}
Now, applying this bound on~\eqref{eq:dp} with the choice $p=\frac{p_0+p_1}{2}$, yields
\begin{align}
h_2(&\alpha p_0+(1-\alpha)p_1)-\alpha h_2(p_0)-(1-\alpha)h_2(p_1)\nonumber\\
&\leq\frac{(p_1-p_0)^2}{4\frac{p_0+p_1}{2}\left(1-\frac{p_0+p_1}{2}\right)}\nonumber\\
&\leq\frac{(p_1-p_0)^2}{2\min\left\{\frac{p_0+p_1}{2},1-\frac{p_0+p_1}{2}\right\}}\nonumber\\
&\leq\frac{(p_1-p_0)^2}{2\min\{p_0,1-p_1\}}
.\label{eq:ub2}
\end{align}
Combining~\eqref{eq:ub1} and~\eqref{eq:ub2} establishes the claim.
\end{IEEEproof}

\medskip

Leveraging this result, it is easy to obtain the following ``sphere-packing'' bound for Bernoulli distributions.

\begin{corollary}
Let $n\geq 1$ be an integer. Given $n+1$ numbers $0\leq p_0 \leq p_1\leq\cdots\leq p_n\leq 1$, there must exist an index $i\in[n]$, such that for any $\alpha\in[0,1]$ we have
\begin{dcalign}
    h_2(\alpha p_{i-1}+(1-\alpha)p_i)-\alpha h_2(p_{i-1})\doubleOnly{&}-(1-\alpha)h_2(p_i)\nonumber\\
    \doubleOnly{&}\leq \frac{8}{(n+1)^2}.\label{eq:spnbound}
\end{dcalign}
\label{cor:twoindices}
\end{corollary}

\begin{IEEEproof}
Let $m=\lfloor\frac{n+1}{2}\rfloor$, and assume $p_{m}\leq 1/2$. We will deal with the case $p_{m}> 1/2$ later. Set $\alpha\in[0,1]$ and let
\begin{align}
    \Delta\triangleq \min_{i \in [n]} \ &h_2(\alpha p_{i-1}+(1-\alpha)p_i)\nonumber\\
&-\alpha h_2(p_{i-1})-(1-\alpha)h_2(p_i).\label{eq:deltadef}
\end{align}
We will show that we must have that $p_m\geq \frac{\Delta(m+1)^2}{4}$, and together with $p_m\leq 1/2$, this will imply that $\Delta\leq \frac{2}{(m+1)^2}$.

By Lemma~\ref{lem:entconcbound}, we have that
\begin{align}
\Delta\leq \min_{i\in[n]}\min\left\{p_i-p_{i-1},\; \frac{(p_i-p_{i-1})^2}{2\min\{p_{i-1},1-p_i\}}\right\}.
\end{align}
In particular, this implies that for any $i\in[n]$ we have
\begin{align}
p_i\geq \max\left\{p_{i-1}+\Delta,\; p_{i-1}+\sqrt{2\Delta \min\{p_{i-1},1-p_i\}} \right\}.\label{eq:pirecursion}
\end{align}
As $p_0\geq 0$,~\eqref{eq:pirecursion} shows that $p_1\geq\Delta$. Since $p_i\leq 1/2$ for all $i\leq m$, we have that $\min\{p_{i-1},1-p_i\}=p_{i-1}$ in this range.
Thus, for all $2\leq i\leq m$, the second term in~\eqref{eq:pirecursion} dominates the maximum, and
\begin{align}
p_{i}\geq p_{i-1}+\sqrt{2\Delta p_{i-1}}.
\end{align}
Defining $\tilde{p}_i=\frac{p_{i}}{\Delta}$, we obtain the recursion:
\begin{align}
\tilde{p}_1&\geq 1\\
\tilde{p}_i&\geq \tilde{p}_{i-1}+\sqrt{2\tilde{p}_{i-1}}, \ 2\leq i\leq m.
\end{align}
It is readily verified by induction that $\tilde{p}_m\geq \frac{(m+1)^2}{4}$, which implies that $p_m\geq \Delta\frac{(m+1)^2}{4}$, as claimed. Combining this with our assumption that $p_m \leq 1/2$ and noting that $m+1 > (n+1)/2$, we establish~\eqref{eq:spnbound}.

If $p_m> 1/2$, we replace each $p_i$ with $\bar{p}_i=1-p_i$, such that now $\bar{p}_m< 1/2$. Now, reorder (reverse) the probabilities such that $0\leq \bar{p}_1\leq\ldots\leq \bar{p}_n$. We are guaranteed that after reordering $\bar{p}_{m'}\leq 1/2$, where $m'=n+1-m = \left\lceil \frac{n+1}{2} \right\rceil \geq \left\lfloor\frac{n+1}{2}\right\rfloor$.
Since $h_2(p)=h_2(1-p)$, we still have
\begin{dcalign}
    \Delta=\min_{i \in [n]} \; \doubleOnly{&}h_2(\alpha \bar{p}_{i-1}+(1-\alpha)\bar{p}_i)\doubleOnly{\nonumber}\\
    \doubleOnly{&}-\alpha h_2(\bar{p}_{i-1})-(1-\alpha)h_2(\bar{p}_i).\label{eq:deltadef2}
\end{dcalign}
Thus, repeating the same argument, we must have that $\Delta \leq \frac{2}{(m'+1)^2}$. Combining this with the fact that $m'+1 > (n+1)/2$, we establish~\eqref{eq:spnbound}.
\label{prop:costofmerge}
\end{IEEEproof}

\section{Upgradation for Binary $\mathcal{X}$}
\label{subsec:binup}

We define a procedure that constructs from a joint distribution $P_{X,Y}$ on $\mathcal{X}\times\mathcal{Y}$ with $|\mathcal{X}|=2$, a distribution $P_{X,Z,Y}$ on $\mathcal{X}\times\mathcal{Z}\times \mathcal{Y}$ with $|\mathcal{Z}|=|\mathcal{Y}|-1$, such that $\sum_{z\in\mathcal{Z}}P_{X,Z,Y}(x,z,y)=P_{X,Y}(x,y)$, and $X-Z-Y$ form a Markov chain in this order.

\begin{definition}[Splitting $i$th symbol of $\mathcal{Y}$]
	Let $(X,Y)\sim P_{X,Y}$ be random variables in $\mathcal{X}\times\mathcal{Y}$, where $\mathcal{X}=\{0,1\}$ and $\mathcal{Y}=\{1,\ldots,|\mathcal{Y}|\}$. Let $p_i=\Prob(X=0|Y=i)$ and assume without loss of generality that $p_1\leq\cdots\leq p_{|\m{Y}|}$. For any $i\in\{2,\ldots,|\m{Y}|-1\}$ let $\alpha_i\in[0,1]$ satisfy $\alpha_i p_{i-1}+(1-\alpha_i)p_{i+1}=p_i$. The joint distribution $P^{i}_{Y,Z,X}=P_Y P^{i}_{Z|Y}P^{i}_{X|Z}$ on $\mathcal{Y}\times\left(\mathcal{Y}\setminus\{i\}\right)\times\mathcal{X}$ corresponding to splitting the $i$th symbol of $\mathcal{Y}$ is defined as
	\begin{align}
        \label{eq:upgradingSplit}
	P_{Z|Y}^i(z|y)=\begin{cases}
	1 & y\neq i,z=y\\
	\alpha_i  & y=i,z=i-1\\
	1-\alpha_i  & y=i,z=i+1\\
	0  & \text{otherwise,}
	\end{cases}
	\end{align}
	and $P_{X|Z}^i(0|z)=p_z$ for all $z\in\m{Y}\setminus\{i\}$.
\end{definition}

\begin{proposition}[Cost of split]
Under the distribution $P^{i}_{Y,Z,X}$, we have that $X-Z-Y$ form a Markov chain in this order, $P^{i}_{X,Y}=P_{X,Y}$, and 
	\begin{dcalign}
        I(X;Z)-I(X;Y)=P_Y\doubleOnly{&}(i)\bigg[h_2(\alpha_i p_{i-1}+(1-\alpha_i)p_{i+1})\nonumber\\
    -\doubleOnly{&}\alpha_i h_2(p_{i-1})-(1-\alpha_i)h_2(p_{i+1})\bigg].\nonumber
	\end{dcalign}
	\label{prop:costofsplit}
\end{proposition}

\begin{IEEEproof}
Clearly, $Y-Z-X$ form a Markov chain in this order under $P^{i}_{Y,Z,X}$, and consequently, so do $X-Z-Y$. By definition $P^{i}_Y=P_Y$, and therefore, to prove that $P^{i}_{X,Y}=P_{X,Y}$, it suffices to show that $P^{i}_{X|Y}=P_{X|Y}$. To that end, write
\begin{align}
P^{i}_{X|Y}(0|y)&=\sum_{z\in\m{Y}\setminus\{i\}}P^{i}_{X|Z}(0|z)P^{i}_{Z|Y}(z|y)\nonumber\\
&=\begin{cases}
p_y & y\neq i\\
\alpha_i p_{i-1}+(1-\alpha_i) p_{i+1} & y=i
\end{cases}\nonumber\\
&=P_{X|Y}(0|y),\nonumber
\end{align}
as $\alpha_i p_{i-1}+(1-\alpha_i) p_{i+1}=p_i$ by definition of $\alpha_i$. 
For the mutual information gap, we first have to compute $P^{i}_{Z,X}=P^{i}_Z P^{i}_{X|Z}$. To this end, write
\begin{align}
P^{i}_{Z}(z)&=\sum_{y\in\m{Y}}P_Y(y)P^{i}_{Z|Y}(z|y)\nonumber\\
&=\begin{cases}
P_{Y}(z) & z\notin\{i-1,i+1\}\\
P_{Y}(i-1)+\alpha_i P_Y(i) & z=i-1\\
P_Y(i+1)+(1-\alpha_i)P_Y(i) & z=i+1
\end{cases},
\end{align}
and recall that $P_{X|Z}^i(0|z)=p_z$.
We can now write
\begin{align}
&I(X;Z)-I(X;Y)=H(X|Y)-H(X|Z)\nonumber\\
&=\sum_{y\in\m{Y}} P_Y(y)h_2(p_y)-\sum_{z\in\m{Y}\setminus\{i\}}P_Z(z)h_2(p_z)\nonumber\\
&=P_Y(i-1)h_2(p_{i-1})+P_Y(i)h_2(p_{i})+P_Y(i+1)h_2(p_{i+1})\nonumber\\
&-\left(P_{Y}(i-1)+\alpha_i P_Y(i)\right)h_2(p_{i-1})\nonumber\\
&-\left(P_Y(i+1)+(1-\alpha_i)P_Y(i)\right)h_2(p_{i+1}) \nonumber\\
&=P_Y(i)h_2(p_{i})-\alpha_i P_{Y}(i)h_2(p_{i-1})-(1-\alpha_i)P_{Y}(i)h_2(p_{i+1})\nonumber\\
&=P_Y(i)\bigg[h_2(\alpha p_{i-1}+(1-\alpha_i)p_{i+1})\doubleOnly{\nonumber\\
& \ \ \ \ \ \ \ \ \ \ \ } -\alpha_i h_2(p_{i-1})-(1-\alpha_i)h_2(p_{i+1})\bigg],\nonumber
\end{align}
as claimed.
\end{IEEEproof}

\begin{theo}
	Let $(X,Y)\sim P_{X,Y}$ be random variables in $\mathcal{X}\times\mathcal{Y}$, where $\mathcal{X}=\{0,1\}$ and $\mathcal{Y}=\{1,\ldots,|\mathcal{Y}|\}$. Let $p_i=\Prob(X=0|Y=i)$ and assume without loss of generality that $p_1\leq\cdots\leq p_{|\mathcal{Y}|}$. Then, there exists $i\in\{2,\ldots,|\mathcal{Y}|-1\}$ such that under the distribution $P^{i}_{Y,Z,X}$, defined in Proposition~\ref{prop:costofsplit}, we have
	\begin{align}
	I(X;Z)-I(X;Y)\leq\frac{256}{|\mathcal{Y}|^3}.\label{eq:onesplit}
	\end{align}
	\label{thm:onestepsplit}
\end{theo}

\begin{IEEEproof}
Without loss of generality we may assume $|\mathcal{Y}|\geq 8$, as otherwise the right hand side of~\eqref{eq:onesplit} is greater than $\log{2}$, and the statement holds trivially. Let 
\begin{align}
\mathcal{Y}_{\text{small}}\triangleq \left\{y\in\mathcal{Y} \ : \ P_{Y}(y)\leq\frac{2}{|\mathcal{Y}|}  \right\},\label{eq:Ysmalldef}
\end{align}
which implies $|\mathcal{Y}_{\text{small}}|\geq\frac{|\mathcal{Y}|}{2}$, and let $\mathcal{Y}_{\text{punctured}}$ be the set obtained by removing every other element in $\mathcal{Y}_{\text{small}}$, starting from the second. We get that between any two elements of $\mathcal{Y}_{\text{punctured}}$ lies at least one element of $\mathcal{Y}_{\text{small}}$, and that $|\mathcal{Y}_{\text{punctured}}|\geq\frac{{|\mathcal{Y}|}}{4}$. Furthermore, by Corollary~\ref{cor:twoindices}, we must have two indices $j,k\in\mathcal{Y}_{\text{punctured}}$, $j>k+1$, such that for any $\alpha\in[0,1]$
\begin{dcalign}
    h_2(\alpha p_{k}+(1-\alpha)p_j)-\alpha h_2(p_{k})-(1-\alpha)h_2(p_j)\doubleOnly{&}\leq \frac{8}{\left(\frac{|\mathcal{Y}|}{4}\right)^2}\nonumber\\
\doubleOnly{&}=\frac{128}{|\mathcal{Y}|^2}.\nonumber
\end{dcalign}
Thus, we must have some $i\in\m{Y}_{\text{small}}$ satisfying $k<i<j$, for which
\begin{dcalign}
    h_2(\alpha_i p_{i-1}+(1-\alpha_i)p_{i+1})\doubleOnly{&}-\alpha_i h_2(p_{i-1})-(1-\alpha_i)h_2(p_{i+1})\nonumber\\
\doubleOnly{&}\leq \frac{128}{|\mathcal{Y}|^2},\nonumber
\end{dcalign}
which follows since\footnote{To see this fix $\alpha$, and denote the LHS of (\ref{eq:h2inequality}) by $g(p',q')$. Recalling that $0 \leq p' \leq q' \leq 1$, it suffices to prove that $g(p',q')$ is non-decreasing if we enlarge $q'$ or reduce $p'$. This is indeed true, and can be seen by considering the partial derivatives of $g(p',q')$ with respect to $p'$ and $q'$, and noting that $x/(1-x)$ is increasing in $x$, for $0 \leq x < 1$.} for $0\leq p\leq p'\leq q'\leq q\leq 1$,
\begin{dcalign}
    \label{eq:h2inequality}
    h_2(\alpha p'\doubleOnly{&}+(1-\alpha)q')-\alpha h_2(p')-(1-\alpha)h_2(q') \\
    \doubleOnly{&}\leq h_2(\alpha p+(1-\alpha)q)-\alpha h_2(p)-(1-\alpha)h_2(q) \doubleOnly{\nonumber} \; .
\end{dcalign}

By Proposition~\ref{prop:costofsplit} and~\eqref{eq:Ysmalldef}, for this $i$ we have that
\begin{align}
I(X;Z)-I(X;Y)\leq \frac{256}{|\mathcal{Y}|^3},\nonumber
\end{align}
under $P^{i}_{YZX}$, as claimed.
\end{IEEEproof}

We are now ready to prove Theorem~\ref{thm:binarysplit}.

\begin{IEEEproof}[Proof of Theorem~\ref{thm:binarysplit}]
	Construct $Z$ in a greedy fashion as follows. Find an index $i$ such that $P^{i}_{Y,Z,X}$ satisfies~\eqref{eq:onesplit}, and then replace $Y\leftarrow Z$, and repeat. Stop when $|\m{Y}|=L$. From Theorem~\ref{thm:onestepsplit} we obtain
	\begin{align}
	I(X;Z)-I(X;Y)&\leq\sum_{\ell=L+1}^{|\m{Y}|}\frac{256}{\ell^3}\nonumber\\
	&<256\int_{t=L}^{\infty}\frac{1}{t^3}dt\nonumber\\
		&=\frac{128}{L^2},\nonumber
	\end{align}
	as claimed.
\end{IEEEproof}

\section{Degradation for Binary $\mathcal{X}$}
\label{subsec:bindown}

For a joint distribution $P_{X,Y}$ on $\mathcal{X}\times\mathcal{Y}$, where $|\mathcal{X}|=2$, we first define a quantizer $f:\mathcal{Y}\to \{1,\ldots, |\mathcal{Y}|-1\}$, with the property that $I(X;f(Y))$ is close to $I(X;Y)$. Then, we apply this quantizer sequentially until the cardinality is reduced to $L$.

\begin{definition}[Merging symbols $i$ and $j$]
Let $(X,Y)\sim P_{X,Y}$ be random variables in $\mathcal{X}\times\mathcal{Y}$, where $\mathcal{X}=\{0,1\}$ and $\mathcal{Y}=\{1,\ldots,|\mathcal{Y}|\}$. The function that merges the symbols $i\neq j\in\mathcal{Y}$ to the symbol $y'$, and does not change the rest of the symbols is defined as $f_{ij}:\mathcal{Y}\to\mathcal{Y}\setminus\{i,j\}\cup y'$, where $y'\notin\mathcal{Y}$.  Namely, $f_{ij}(i)=f_{ij}(j)=y'$ and $f(y)=y$ for all $y\notin\{i,j\}$.
\end{definition}

\medskip

\begin{proposition}[Cost of merge]
Let $p_i=\Prob(X=0|Y=i)$, and $\alpha_{ij}=\frac{P_Y(i)}{P_{Y}(i)+P_Y(j)}$. Then,
\begin{align}
&I(X;Y)-I(X;f_{ij}(Y))=\left(P_Y(i)+P_Y(j)\right)\nonumber\\
&\cdot\left[h_2(\alpha_{ij} p_{i}+(1-\alpha_{ij})p_j)-\alpha_{ij} h_2(p_{i})-(1-\alpha_{ij})h_2(p_j)\right].\nonumber
\end{align} 
\label{prop:costofmerg}
\end{proposition}

\begin{IEEEproof}
Let $\tilde{Y}=f_{ij}(Y)$. We clearly have that $P_{\tilde{Y}}(y)=P_Y(y)$ and $\Prob(X=0|\tilde{Y}=y)=p_i$ for any $y\in\m{Y}\setminus\{i,j\}$, while for $y'$ we have  $P_{\tilde{Y}}(y')=P_Y(i)+P_Y(j)$ and  $\Prob(X=0|\tilde{Y}=y')=\frac{P_Y(i)p_i+P_Y(j)p_j}{P_Y(i)+P_Y(j)}=\alpha_{ij} p_i+(1-\alpha_{ij})p_j$.
It therefore follows that
\begin{align}
&I(X;Y)-I(X;f(Y))=H(X|f(Y))-H(X|Y)\nonumber\\
&=\left(P_Y(i)+P_Y(j)\right)h_2\left(\alpha_{ij} p_i+(1-\alpha_{ij})p_j\right)\nonumber\\
&+\sum_{y\in\m{Y}\setminus\{i,j\}}P_{Y}(y)h_2(p_y)\nonumber\\
&-P_Y(i)h_2(p_i)-P_Y(j)h_2(p_j)-\sum_{y\in\m{Y}\setminus\{i,j\}}P_{Y}(y)h_2(p_y)\nonumber\\
&=\left(P_Y(i)+P_Y(j)\right)h_2\left(\alpha_{ij} p_i+(1-\alpha_{ij})p_j\right)\nonumber\\
&-P_Y(i)h_2(p_i)-P_Y(j)h_2(p_j),\nonumber
\end{align}
and the result follows by definition of $\alpha_{ij}$.
\end{IEEEproof}

\medskip

The following theorem shows that if $|\mathcal{Y}|$ is large, we can always find two symbols such that merging them would not significantly decrease the mutual information.

\begin{remark}
The merging operation is in fact a quantizer of $Y$ to $|\m{Y}|-1$ levels. It is known~\cite{bdk92,wcx2000,KurkoskiYagi:14p} that an optimal quantizer $f:\m{Y}\to [M]$, in terms of maximizing $I(X;f(Y)$, can be associated with a partition of the interval $[0,1]$ to $M$ disjoint intervals $\mathcal{I}_1,\ldots,\mathcal{I}_M$, such that $f(y)=m$ iff $\Prob(X=0|Y=y)\in\mathcal{I}_m$. Thus, if we relabel $\m{Y}=\{1,\ldots,|\m{Y}|\}$ such that $p_1\leq \cdots\leq p_{|\m{Y}|}$, we have that the symbols with the smallest cost of merge are adjacent. It therefore suffices to restrict the search to $i\in\{1,\ldots,|\m{Y}|-1\}$ and $j=i+1$.
\label{rem:adjacentmerge}
\end{remark}

\begin{theo}
Let $(X,Y)\sim P_{X,Y}$ be random variables in $\mathcal{X}\times\mathcal{Y}$, where $\mathcal{X}=\{0,1\}$, $\mathcal{Y}=\{1,\ldots,|\mathcal{Y}|\}$, and let $f_{ij}$ be as defined in Proposition~\ref{prop:costofmerg}. Then, there exists $i\neq j\in\mathcal{Y}$ such that
\begin{align}
I(X;Y)-I(X;f_{ij}(Y))\leq\frac{128}{|\mathcal{Y}|^3}.\label{eq:onestepUB}
\end{align}
\label{thm:onestep}
\end{theo}

\begin{IEEEproof}
Without loss of generality we may assume $|\mathcal{Y}|\geq4$, as otherwise the right hand side of~\eqref{eq:onestepUB} is greater than $\log{2}$ and the statement holds trivially. Let $\mathcal{Y}_{\text{small}}$ be as in~\eqref{eq:Ysmalldef}, and recall that
$|\mathcal{Y}_{\text{small}}|\geq\frac{|\mathcal{Y}|}{2}$. By Corollary~\ref{cor:twoindices}, there must exist $i\neq j\in\mathcal{Y}_{\text{small}}\subset\mathcal{Y}$ for which
\begin{dcalign}
    h_2(\alpha_{ij} p_{i}\doubleOnly{&}+(1-\alpha_{ij})p_j)-\alpha_{ij} h_2(p_{i})-(1-\alpha_{ij})h_2(p_j)\doubleOnly{\nonumber}\\
\doubleOnly{&}\leq \frac{8}{|\mathcal{Y}_{\text{small}}|^2}\leq\frac{32}{|\mathcal{Y}|^2}.\label{eq:entcost}
\end{dcalign}
Furthermore, as $i,j\in\m{Y}_{\text{small}}$, we have that 
\begin{align}
P_{Y}(i)+P_Y(j)\leq\frac{4}{|\mathcal{Y}|}.\label{eq:sumprob}
\end{align} 
The claim now immediately follows from substituting~\eqref{eq:entcost} and~\eqref{eq:sumprob} into  Proposition~\ref{prop:costofmerg}.
\end{IEEEproof}

\begin{theo}
Let $(X,Y)\sim P_{X,Y}$ be random variables in $\mathcal{X}\times\mathcal{Y}$, where $|\mathcal{X}|=2$ and $\mathcal{Y}$ is discrete. For any integer $L$, the greedy-merge algorithm finds a quantizer $f:\m{Y}\to \{1,\ldots,L\}$ such that
\begin{align}
I(X;Y)-I(X;f(Y))\leq \frac{64}{L^2}.
\end{align}
\label{thm:greedymergebin}
\end{theo}

\begin{IEEEproof}
Construct $f$ in a greedy fashion as follows. Merge the two symbols of $\mathcal{Y}$ for which the loss in mutual information due to merging is smallest, and repeat this until $|\mathcal{Y}|=L$. From Theorem~\ref{thm:onestep} we obtain
\begin{align}
I(X;Y)-I(X;f(Y))&\leq\sum_{\ell=L+1}^{|\mathcal{Y}|}\frac{128}{\ell^3}\nonumber\\
&<128\int_{t=L}^{\infty}\frac{1}{t^3}dt\nonumber\\
&= \frac{64}{L^2},
\end{align}
as claimed.
\end{IEEEproof}

\section{Degradation for General $\mathcal{X}$}
\label{subsec:gendown}

Repeating the proof of~\cite[Proposition 7]{bnop18}, with the improved performance guarantees of the greedy merge algorithm for binary $\mathcal{X}$, as stated in Theorem~\ref{thm:greedymergebin}, yields the bound~\eqref{eq:degradeonehotcost} from Theorem~\ref{thm:greedymergegen}. For completeness, we bring the proof below.

\begin{IEEEproof}[Proof of~\eqref{eq:degradeonehotcost} from  Theorem~\ref{thm:greedymergegen}]
The case $q=2$ is covered by Theorem~\ref{thm:greedymergebin}.
	
Now let $q > 2$, 
and without loss of generality assume $\mathcal{X}=\{1,2,\ldots,q\}$. Define $X_i \triangleq \Ind_{\{X = i\}}$, for $i=1,2,\ldots,q-1$. Then, 
	\begin{dcalign}
        I(X;Y) \doubleOnly{&}= I(X_1,\ldots,X_{q-1};Y) \doubleOnly{\nonumber} \\
        \doubleOnly{&}= \sum_{i=1}^{q-1} I(X_i;Y|X_1^{i-1}=0_1^{i-1})\Prob(X_1^{i-1}=0_1^{i-1}) \label{eq:additiveone}
	\end{dcalign}
	where $X_1^{i-1}=0_1^{i-1}$ denotes the event $X_1  = \cdots = X_{i-1} = 0$. 
	
	Let $f(y)$ be an $M$-level quantizer, with $M\leq L$, of the form $f(y) = (f_1(y),\ldots,f_{q-1}(y))$. Then,

    \singleOnly{
    \begin{IEEEeqnarray}{rCl}
        I(X;f(Y)) &=&\sum_{i=1}^{q-1} I(X_i;f(Y)|X_1^{i-1}=0_1^{i-1})\Prob(X_1^{i-1}=0_1^{i-1})\nonumber\\
                  &\ge& \sum_{i=1}^{q-1} I(X_i;f_i(Y)|X_1^{i-1}=0_1^{i-1})\Prob(X_1^{i-1}=0_1^{i-1}). \label{eq:additivetwo}
\end{IEEEeqnarray}
}
\doubleOnly{
    \begin{IEEEeqnarray}{l}
        I(X;f(Y)) \nonumber \\
        =\sum_{i=1}^{q-1} I(X_i;f(Y)|X_1^{i-1}=0_1^{i-1})\Prob(X_1^{i-1}=0_1^{i-1})\nonumber\\
                  \ge \sum_{i=1}^{q-1} I(X_i;f_i(Y)|X_1^{i-1}=0_1^{i-1})\Prob(X_1^{i-1}=0_1^{i-1}). \label{eq:additivetwo}
\end{IEEEeqnarray}
}

	Thus, combining~\eqref{eq:additiveone} and~\eqref{eq:additivetwo}, gives
	\begin{dcalign}
        \doubleOnly{&}I(X;Y)-I(X;f(Y))\le \sum_{i=1}^{q-1} \big(I(X_i;Y|X_1^{i-1}=0_1^{i-1})\doubleOnly{\nonumber} \\
        \doubleOnly{&}   - I(X_i;f_i(Y)|X_1^{i-1}=0_1^{i-1})\big)\Prob(X_1^{i-1}=0_1^{i-1}).
	\end{dcalign}
	It follows from Theorem~\ref{thm:greedymergebin} that by taking the support sizes as $|f_i(y)| =\Lambda$, where $\Lambda=\left\lfloor L^{1/(q-1)}\right\rfloor$,  for all $1 \le i \le q-1$, that the quantizers
	 $f_1(y),\ldots,f_{q-1}(y)$ returned by the greedy-merge algorithm satisfy
	\begin{align*}
	I(X_i;Y|X_1^{i-1}=0_1^{i-1})-I(X_i;f_i(Y)|X_1^{i-1}=0_1^{i-1}) \le \frac{64}{\Lambda^{2}}. 
	\end{align*}
	Consequently, with this choice, we obtain
	\begin{align}
	I(X;Y)-I(X;f(Y)) &\le 64\cdot \Lambda^{-2} \sum_{i=1}^{q-1}\Prob(X_1^{i-1}=0)\nonumber\\
	&\le 64(q-1) \cdot \left\lfloor L^{1/(q-1)}\right\rfloor^{-2},
	\end{align}
	as desired.
\end{IEEEproof}

\end{appendices}

\twobibs{
\bibliographystyle{IEEEtran}
\bibliography{mybib.bib}
}
{
\ifdefined\bibstar\else\newcommand{\bibstar}[1]{}\fi

}
\end{document}

%% file: numerical_example_upgrade.tex
%
%
\newcommand{\linewidthpt}{2pt}
\definecolor{mycolor1}{rgb}{0.00000,1.00000,1.00000}%
\begin{tikzpicture}[scale=0.5]

\begin{axis}[%
width=5.888in,
height=4.477in,
at={(0.988in,0.604in)},
scale only axis,
xmode=log,
xmin=5,
xmax=512,
xminorticks=true,
xlabel style={font=\color{white!15!black}},
xlabel={$L$},
ymode=log,
ymin=1e-05,
ymax=100,
yminorticks=true,
ylabel style={font=\color{white!15!black}},
ylabel={$\Delta I$},
axis background/.style={fill=white},
xmajorgrids,
xminorgrids,
ymajorgrids,
yminorgrids,
legend style={at={(0.03,0.03)}, anchor=south west, legend cell align=left, align=left, draw=white!15!black}
]

\addplot [color=red, line width=\linewidthpt]
  table[row sep=crcr]{%
10 41.036658940841626\\
30 14.773197218702986\\
50 7.53734551974642\\
70 5.7707801635558535\\
90 4.559628771204625\\
110 3.6932993046757465\\
130 3.0523134749386336\\
150 2.5647911838026016\\
170 2.1853842039501457\\
190 2.1853842039501457\\
210 1.884336379936605\\
230 1.641466357633665\\
250 1.641466357633665\\
270 1.4426950408889634\\
290 1.277958236912023\\
310 1.277958236912023\\
330 1.1399071928011562\\
350 1.1399071928011562\\
370 1.0230745996331707\\
390 1.0230745996331707\\
410 0.9233248261689366\\
430 0.9233248261689366\\
450 0.8374828355273801\\
470 0.8374828355273801\\
490 0.7630783687346584\\
510 0.7630783687346584\\
530 0.6981662201655475\\
550 0.6981662201655475\\
570 0.6981662201655475\\
590 0.6411977959506504\\
610 0.6411977959506504\\
630 0.5909278887481194\\
650 0.5909278887481194\\
670 0.5909278887481194\\
690 0.5463460509875364\\
710 0.5463460509875364\\
};
\addlegendentry{Upper bound in (\ref{eq:theoDifferenceInEntropies})}

\addplot [color=magenta, dotted, line width=\linewidthpt]
  table[row sep=crcr]{%
10 1.1911059674761177\\
19 0.8335801132704421\\
31 0.6807265223577852\\
46 0.5430726769154299\\
64 0.46638872108812224\\
85 0.40376338372118237\\
109 0.3796114109102581\\
157 0.32827315637360044\\
190 0.29308822976004734\\
205 0.2742722710608759\\
265 0.26158476926892327\\
301 0.2350833104062625\\
352 0.2127840902343595\\
400 0.19803908207446597\\
451 0.1950320332044857\\
520 0.18417689010669824\\
565 0.16633814817666148\\
622 0.15636740538883376\\
730 0.1557770588590326\\
};
\addlegendentry{The upgrading algorithm in \cite{PeregTal:17p}}

\addplot [color=blue, dashed, line width=\linewidthpt]
  table[row sep=crcr]{%
4 1.1986153176675076\\
9 0.3339467289488509\\
15 0.20970220268098694\\
24 0.10821365576745778\\
35 0.06403851238917646\\
48 0.043203785365554914\\
63 0.03501793956766708\\
80 0.02723166723951964\\
99 0.022306289036828852\\
120 0.017893670514815607\\
143 0.014388335917611128\\
168 0.01181536404600414\\
195 0.010111405870623758\\
224 0.00859053398418963\\
255 0.007454469882316461\\
286 0.00668760963681958\\
321 0.0057512248436004665\\
358 0.004953721334463035\\
397 0.004530712619180033\\
438 0.0041718880845820205\\
481 0.003814640314565798\\
525 0.0034985624598051768\\
571 0.0032063741960237646\\
620 0.0029471159788918566\\
671 0.0027018821230606616\\
724 0.002460391668131612\\
};
\addlegendentry{One-hot upgrading}

\addplot [color=black, line width=\linewidthpt]
  table[row sep=crcr]{%
10 0.005740301178291065\\
30 0.001913433726097022\\
50 0.001148060235658213\\
70 0.0008200430254701521\\
90 0.0006378112420323406\\
110 0.000521845561662824\\
130 0.0004415616290993128\\
150 0.0003826867452194044\\
170 0.0003376647751935921\\
190 0.0003021211146468982\\
210 0.00027334767515671744\\
230 0.00024957831209961154\\
250 0.0002296120471316426\\
270 0.00021260374734411355\\
290 0.0001979414199410712\\
310 0.0001851710057513247\\
330 0.0001739485205542747\\
350 0.00016400860509403043\\
370 0.00015514327508894772\\
390 0.0001471872096997709\\
410 0.0001400073458119772\\
430 0.0001334953762393271\\
450 0.00012756224840646812\\
470 0.00012213406762321414\\
490 0.00011714900363859318\\
510 0.00011255492506453068\\
530 0.00010830756940171822\\
550 0.00010436911233256483\\
570 0.00010070703821563274\\
590 9.729324031001805e-05\\
610 9.410329800477157e-05\\
630 9.11158917189058e-05\\
650 8.831232581986254e-05\\
670 8.567613698941888e-05\\
690 8.319277069987051e-05\\
710 8.08493123702967e-05\\
};
\addlegendentry{Lower bound on Upgrading Cost, \cite[equation (43)]{KartowskyTal:17p}}

%
\end{axis}
\end{tikzpicture}%

%% file: numerical_example_degrade.tex
%
%
\newcommand{\linewidthpt}{2pt}
\definecolor{mycolor1}{rgb}{0.00000,1.00000,1.00000}%
\begin{tikzpicture}[scale=0.5]

\begin{axis}[%
width=5.888in,
height=4.477in,
at={(0.988in,0.604in)},
scale only axis,
xmode=log,
xmin=5,
xmax=512,
xminorticks=true,
xlabel style={font=\color{white!15!black}},
xlabel={$L$},
ymode=log,
ymin=1e-05,
ymax=100,
yminorticks=true,
ylabel style={font=\color{white!15!black}},
ylabel={$\Delta I$},
axis background/.style={fill=white},
xmajorgrids,
xminorgrids,
ymajorgrids,
yminorgrids,
legend style={at={(0.03,0.03)}, anchor=south west, legend cell align=left, align=left, draw=white!15!black}
]

\addplot [color=red, line width=\linewidthpt]
  table[row sep=crcr]{%
10 20.518329470420813\\
30 7.386598609351493\\
50 3.76867275987321\\
70 2.8853900817779268\\
90 2.2798143856023123\\
110 1.8466496523378733\\
130 1.5261567374693168\\
150 1.2823955919013008\\
170 1.0926921019750728\\
190 1.0926921019750728\\
210 0.9421681899683025\\
230 0.8207331788168325\\
250 0.8207331788168325\\
270 0.7213475204444817\\
290 0.6389791184560115\\
310 0.6389791184560115\\
330 0.5699535964005781\\
350 0.5699535964005781\\
370 0.5115372998165854\\
390 0.5115372998165854\\
410 0.4616624130844683\\
430 0.4616624130844683\\
450 0.41874141776369006\\
470 0.41874141776369006\\
490 0.3815391843673292\\
510 0.3815391843673292\\
530 0.34908311008277376\\
550 0.34908311008277376\\
570 0.34908311008277376\\
590 0.3205988979753252\\
610 0.3205988979753252\\
630 0.2954639443740597\\
650 0.2954639443740597\\
670 0.2954639443740597\\
690 0.2731730254937682\\
710 0.2731730254937682\\
};
\addlegendentry{Upper bound in (\ref{eq:degradeonehotcost})}

\addplot [color=magenta, dotted, line width=\linewidthpt]
  table[row sep=crcr]{%
7 0.10692211946977181\\
16 0.09294578134585718\\
28 0.0404136345649877\\
43 0.04097813622685065\\
67 0.021935806170019445\\
94 0.023810325270747335\\
118 0.014198756812199953\\
151 0.015426367207096003\\
181 0.010348696238419164\\
232 0.011088002487698922\\
274 0.007875853331295923\\
310 0.008327970531109674\\
358 0.006136698195918333\\
406 0.006652633063150004\\
469 0.004857452760390624\\
550 0.005358734494996575\\
598 0.004013271646707395\\
667 0.004428232270085708\\
709 0.0033683008725582386\\
};
\addlegendentry{The degrading algorithm in \cite{TalSharovVardy:12c}}

\addplot [color=blue, dashed, line width=\linewidthpt]
  table[row sep=crcr]{%
4 0.13562284075062503\\
9 0.07668756950517297\\
16 0.044095641898525884\\
25 0.029233448076286583\\
36 0.021221230560787063\\
49 0.0157424189940345\\
64 0.011893711956363795\\
81 0.009459376804689112\\
100 0.007948607236509142\\
121 0.0068008289107452935\\
144 0.0057422281346783954\\
169 0.004945791234432173\\
196 0.0042423469663355196\\
225 0.0036560767103090974\\
256 0.0031572532947694576\\
289 0.0027952454996553744\\
324 0.002480883191033234\\
361 0.002283001362397119\\
400 0.0021041899959848287\\
441 0.001942225409153675\\
484 0.0017970639165671987\\
529 0.001659243869009508\\
575 0.00154042553451883\\
623 0.0014295855024299975\\
674 0.0013299294222246516\\
727 0.0012304857244689327\\
};
\addlegendentry{One-hot degrading}

\addplot [color=black, line width=\linewidthpt]
  table[row sep=crcr]{%
10 0.005740301178291065\\
30 0.001913433726097022\\
50 0.001148060235658213\\
70 0.0008200430254701521\\
90 0.0006378112420323405\\
110 0.000521845561662824\\
130 0.0004415616290993128\\
150 0.0003826867452194044\\
170 0.00033766477519359204\\
190 0.0003021211146468982\\
210 0.0002733476751567174\\
230 0.0002495783120996115\\
250 0.0002296120471316426\\
270 0.00021260374734411355\\
290 0.0001979414199410712\\
310 0.0001851710057513247\\
330 0.0001739485205542747\\
350 0.00016400860509403046\\
370 0.00015514327508894772\\
390 0.0001471872096997709\\
410 0.0001400073458119772\\
430 0.0001334953762393271\\
450 0.00012756224840646812\\
470 0.00012213406762321414\\
490 0.00011714900363859316\\
510 0.00011255492506453068\\
530 0.0001083075694017182\\
550 0.00010436911233256482\\
570 0.00010070703821563271\\
590 9.729324031001805e-05\\
610 9.410329800477155e-05\\
630 9.111589171890579e-05\\
650 8.831232581986254e-05\\
670 8.567613698941888e-05\\
690 8.31927706998705e-05\\
710 8.08493123702967e-05\\
};
\addlegendentry{Lower bound on Degrading Cost, \cite[equation (3)]{Tal:17p}}

%
\end{axis}
\end{tikzpicture}%